\documentclass[10pt,letterpaper]{article}
\usepackage[top=0.85in,left=2.75in,footskip=0.75in]{geometry}

\usepackage{changepage}

\usepackage[utf8]{inputenc}

\usepackage{textcomp,marvosym}

\usepackage{fixltx2e}

\usepackage{amsmath,amssymb}

\usepackage{cite}

\usepackage{nameref,hyperref}

\usepackage[right]{lineno}

\usepackage{microtype}
\DisableLigatures[f]{encoding = *, family = * }

\usepackage{rotating}

\raggedright
\usepackage[aboveskip=1pt,labelfont=bf,labelsep=period,justification=raggedright,singlelinecheck=off]{caption}

\makeatletter
\renewcommand{\@biblabel}[1]{\quad#1.}
\makeatother

\date{}

\usepackage{lastpage,fancyhdr,graphicx}
\usepackage{epstopdf}
\fancyheadoffset[L]{2.25in}
\fancyfootoffset[L]{2.25in}

\begin{document}
\vspace*{0.35in}

\begin{flushleft}
{\Large
\textbf\newline{Pluripotency, differentiation, and reprogramming: A gene expression dynamics model with epigenetic feedback regulation}
}
\newline
\\
Tadashi Miyamoto\textsuperscript{1},
Chikara Furusawa\textsuperscript{2},
Kunihiko Kaneko\textsuperscript{1, *}
\\
\bigskip
\bf{1} Department of Basic Science, The University of Tokyo, Meguro-ku, Tokyo, Japan
\\
\bf{2} Laboratory for Multiscale Biosystem Dynamics, Quantitative Biology Center, RIKEN, Suita, Osaka, Japan
\\
\bigskip

* kaneko@complex.c.u-tokyo.ac.jp
\end{flushleft}
\section*{Abstract}
Embryonic stem cells exhibit pluripotency: they can differentiate into all types of somatic cells. Pluripotent genes such as \textit{Oct4} and \textit{Nanog} are activated in the pluripotent state, and their expression decreases during cell differentiation. Inversely, expression of differentiation genes such as \textit{Gata6} and \textit{Gata4} is promoted during differentiation. The gene regulatory network controlling the expression of these genes has been described, and slower-scale epigenetic modifications have been uncovered. Although the differentiation of pluripotent stem cells is normally irreversible, reprogramming of cells can be experimentally manipulated to regain pluripotency via overexpression of certain genes. Despite these experimental advances, the dynamics and mechanisms of differentiation and reprogramming are not yet fully understood.
Based on recent experimental findings, we constructed a simple gene regulatory network including pluripotent and differentiation genes, and we demonstrated the existence of pluripotent and differentiated states from the resultant dynamical-systems model. Two differentiation mechanisms, interaction-induced switching from an expression oscillatory state and noise-assisted transition between bistable stationary states, were tested in the model. The former was found to be relevant to the differentiation process. We also introduced variables representing epigenetic modifications, which controlled the threshold for gene expression. By assuming positive feedback between expression levels and the epigenetic variables, we observed differentiation in expression dynamics. Additionally, with numerical reprogramming experiments for differentiated cells, we showed that pluripotency was recovered in cells by imposing overexpression of two pluripotent genes and external factors to control expression of differentiation genes. Interestingly, these factors were consistent with the four Yamanaka factors, \textit{Oct4, Sox2, Klf4}, and \textit{Myc}, which were necessary for the establishment of induced pluripotent cells.
These results, based on a gene regulatory network and expression dynamics, contribute to our wider understanding of pluripotency, differentiation, and reprogramming of cells, and they provide a fresh viewpoint on robustness and control during development.

\section*{Author Summary}
Characterization of pluripotent states, in which cells can both self-renew and differentiate, and the irreversible loss of pluripotency are important research areas in developmental biology. In particular, an understanding of these processes is essential to the reprogramming of cells for biomedical applications, i.e., the experimental recovery of pluripotency in differentiated cells. Based on recent advances in dynamical-systems theory for gene expression, we propose a gene-regulatory-network model consisting of several pluripotent and differentiation genes. Our results show that cellular-state transition to differentiated cell types occurs as the number of cells increases, beginning with the pluripotent state and oscillatory expression of pluripotent genes. Cell-cell signaling mediates the differentiation process with robustness to noise, while epigenetic modifications affecting gene expression dynamics fix the cellular state. These modifications ensure the cellular state to be protected against external perturbation, but they also work as an epigenetic barrier to recovery of pluripotency. We show that overexpression of several genes leads to the reprogramming of cells, consistent with the methods for establishing induced pluripotent stem cells. Our model, which involves the inter-relationship between gene expression dynamics and epigenetic modifications, improves our basic understanding of cell differentiation and reprogramming.


\section*{Introduction}
In multicellular organisms, cells that exhibit stemness during development can both self-renew and differentiate into other cell types. In contrast, differentiated cells lose the ability to further differentiate into other cell types and terminally differentiated cells can only self-renew. Currently, how stemness and the irreversible loss of differentiation potential are characterized by gene expression patterns and dynamics are key questions in developmental biology.

Cells with stemness include embryonic stem cells (ESCs), which are derived from the inner cell mass of a mammalian blastocyst and are pluripotent, i.e., they can differentiate into all the types of somatic cells\cite{EvansKaufman1981,MrtinGR1981}. To maintain pluripotency, pluripotent genes such as \textit{Pou5f1} (also known as \textit{Oct4})\cite{NicholsJ1998,NiwaH2000} and \textit{Nanog}\cite{ChambersI2003,MitsuiK2003} are activated in ESCs. Expression of these genes gradually decreases during cell differentiation, whereas expression of differentiation marker genes increases. Understanding these changes in gene expression patterns over the course of cell differentiation is important for characterizing the loss of pluripotency.

During normal development, the loss of pluripotency is irreversible. However, the recovery of pluripotency in differentiated cells was first achieved by experimental manipulation in plants, and then in \textit{Xenopus laevis} via cloning by Gurdon\cite{GurdonJB1962}. More recently, the overexpression of four genes that are highly expressed in ECSs, \textit{Oct4, Sox2, Klf4}, and \textit{Myc} (now termed Yamanaka factors), has been used to reprogram differentiated cells. Overexpression of these genes leads to cellular-state transition and changes in gene expression patterns, and the transition generates cells known as induced pluripotent stem cells (iPSCs)\cite{TakahashiYamanaka2006}. Previous studies have also uncovered the gene regulatory network (GRN) related to the differentiation and reprogramming of cells \cite{BoyerLA2005,HanJ2011}.

To understand the differentiation process theoretically, Waddington proposed a landscape scenario in which each stable cell-type is represented as a valley and the differentiation process is represented as a ball rolling from the top of a hill down into the valley\cite{Waddington1957}. In this scenario, the reprogramming process works inversely to push the ball to the top of the hill\cite{MohammadHP2010,YamanakaS2009,GrafT2009}.

As a theoretical representation of Waddington's landscape, the dynamical-systems approach has been developed over several decades, pioneered by Kauffman\cite{KauffmanSA1969} and Goodwin\cite{GoodwinBC1963}. In this approach, the cellular state is represented by a set of protein expression levels with temporal changes that are given by GRNs. According to gene expression dynamics, the cellular state is attracted to one of the stable states, which is termed an attractor. Each attractor is assumed to correspond to each cell type.

Indeed, this attractor view has become important for understanding the diversification of cellular states and their robustness. Both theoretical and experimental approaches have been developed to assign each cell-type to one of the multi-stable states\cite{HuangS2011,ShuJ2013,WangJ2011}. In these approaches, a pluripotent state is regarded as a stationary attractor with relatively weak stability, and the loss of pluripotency is the transition by noise to attractors with stronger stability.

An alternative approach investigated how the interplay between intra-cellular dynamics and interaction leads to differentiation and the loss of pluripotency\cite{JTB2001,GotoY2013, Kurths2014, Kurths2013}. Specifically, the pluripotent state is represented by oscillatory states following the expression dynamics of more genes, whereas the loss of pluripotency is represented by the decrease in the degree of expressed genes necessary for oscillatory dynamics. Here, differentiation is triggered by cell-cell interactions, which lead to robustness in developmental paths and the final distribution of cell types \cite{KanekoYomo1994,KanekoYomo1997,JTB2001}. By using several GRNs, cells with oscillatory intracellular gene expression dynamics are found to differentiate into other cell types by cell-cell interactions\cite{SuzukiN2011,GotoY2013, Koseska2010, Ullner2007}. Indeed, the recovery of pluripotency by gene overexpression is a process predicted to facilitate recovery of lost degrees of freedom and oscillation\cite{JTB2001}. However, of the question of whether this theory applies to realistic GRNs has yet to be explored. Despite these earlier studies, pluripotency has not yet been confirmed in a realistic GRN observed in experiments, and the mechanism of reprogramming remains elusive.

Epigenetic modifications such as DNA methylation and histone modification are now also recognized as important in cell differentiation. Epigenetic change solidifies differentiated-cellular states by altering chromatin structure to generate transcriptionally active and inactive regions\cite{ErnstJ2011,MarstrandTT2014}. With epigenetic change, the activity of gene expression is preserved in a process known as epigenetic memory\cite{KimK2010}. Indeed, epigenetic modification is suggested as a barrier to reprogramming\cite{ChenJ2013}. However, the theoretical inter-relationship between expression dynamics and epigenetic modification has yet to be fully explored.

The aim of the present study was three-fold. First, by using GRNs obtained from a previous experimental study, we examined the validity of two differentiation scenarios: 1) oscillation + cell-cell interaction and 2) multistability + noise. Second, to demonstrate that differentiation by gene expression dynamics is solidified by epigenetic modification, we introduced a mathematical model for epigenetic feedback regulation. Third, we investigated how overexpression of some genes leads to reprogramming, i.e., regaining pluripotency from differentiated states (scenario 1) by initializing epigenetic changes.

Below, we have first introduced a simple model extracted from an experimentally observed GRN. This model consists of several genes, including pluripotent and differentiation genes, with mutual activation and inhibition. We then examined the oscillatory dynamics and multistable states scenarios to show that differentiation with the loss of pluripotency progresses from a stem cell state with oscillatory expression through cell-cell interactions. Additionally, the two scenarios were compared with regard to their robustness to noise and the regulation of the ratio of differentiated cells.

We also investigated the epigenetic process by introducing variables that give the threshold for the expression of genes to demonstrate that the cellular state derived from gene expression dynamics is fixed by epigenetic feedback regulation. Differentiation by gene expression is fixed according to these threshold variables; thus, the pluripotent and differentiated states are fixed.

Finally, we investigated reprogramming by temporally imposing overexpression of genes and examining whether the differentiated state is reversed to the pluripotent state. Via overexpression of several genes, epigenetic fixation was relaxed and the expression levels and dynamics of the pluripotent state were recovered. This reprogramming was shown to require the overexpression of several genes, including pluripotent genes, over a sufficient period beyond the time scale of epigenetic fixation. Indeed, by using a model with five relevant genes, we found that four genes corresponding to the Yamanaka factors must be overexpressed for reprogramming to occur. It was also demonstrated that insufficient overexpression of genes, i.e., overexpression of pluripotent genes only, results in partially reprogrammed cells (which, experimentally, are known as pre-iPSCs).

\section*{Construction of GRN model}
In the pluripotent state, cells can proliferate and retain their potentiality for differentiation. The expression of pluripotent genes is necessary for pluripotency, but it is not always sufficient. In the differentiation process, expression of pluripotent genes gradually decreases, while expression of differentiation marker genes increases. These temporal changes are a result of gene-gene regulation, which can be integrated as a GRN consisting of pluripotent and differentiation genes.

\begin{figure}[h]
\includegraphics*[width=9cm, bb=0 0 1863 1722]{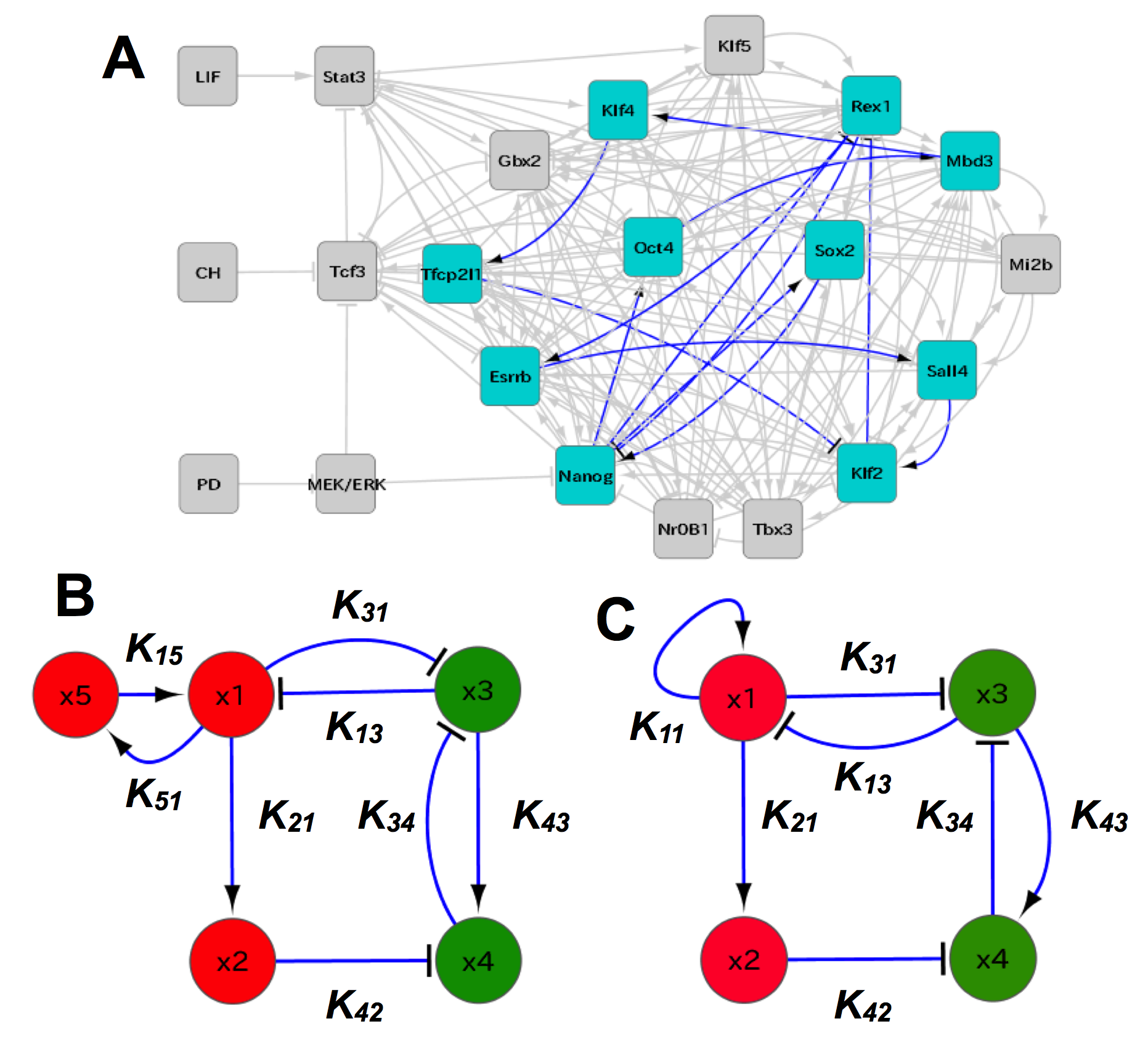}
\caption{{\bf Gene regulatory network.}
A: The GRN is inferred as a core pluripotency network by Dunn et al.\cite{SmithAG2014}. It includes pluripotent transcription factors such as \textit{Oct4} and \textit{Nanog}. In this paper, we first picked up only the eleven genes depicted by cyan nodes, which are considered to be relevant to pluripotency and reprogramming, and include those concerned with Yamanaka four factors, while experimentally confirmed interactions among them as depicted by blue edges are adopted\cite{BoyerLA2005}. We then reduced them to four or five nodes, by compressing a linear chain part A $\rightarrow$ B $\rightarrow$ C to A $\rightarrow$ C, or A $\rightarrow$ B $\dashv$ C to A $\dashv$ C, where $\rightarrow$ represents activation and $\dashv$ inhibition. B: The five-gene simplified model. The regulator from \textit{Oct4} to \textit{Klf2} is compressed into that from $x_2$ to $x_4$, while the regulator from \textit{Rex1} to \textit{Klf2} is compressed into that from $x_3$ to $x_4$, where $x_1$, $x_2$, $x_3$, $x_4$, and $x_5$ correspond to \textit{Nanog, Oct4, Gata6, Gata4}, and \textit{Klf4}, respectively. C: The four-gene model, consisting of two pluripotent (red; $x_1, x_2$) and differentiation (green; $x_3, x_4$) marker genes, in which the positive feedback from $x_5$ to $x_1$ is replaced by auto-regulation. In all diagrams, arrow-headed and T-headed lines represent positive and negative regulation, respectively.}
\label{Fig1}
\end{figure}

Here, we adopted the GRN reported by Dunn et al.\cite{SmithAG2014} (Fig.~\ref{Fig1}) and produced simplified models by compressing some paths and genes while maintaining the structure of the GRN (see Models).

\section*{Results}
\subsection*{Single cell dynamics}
\begin{figure}[h]
\includegraphics*[width=15cm, bb=0 0 4000 2000]{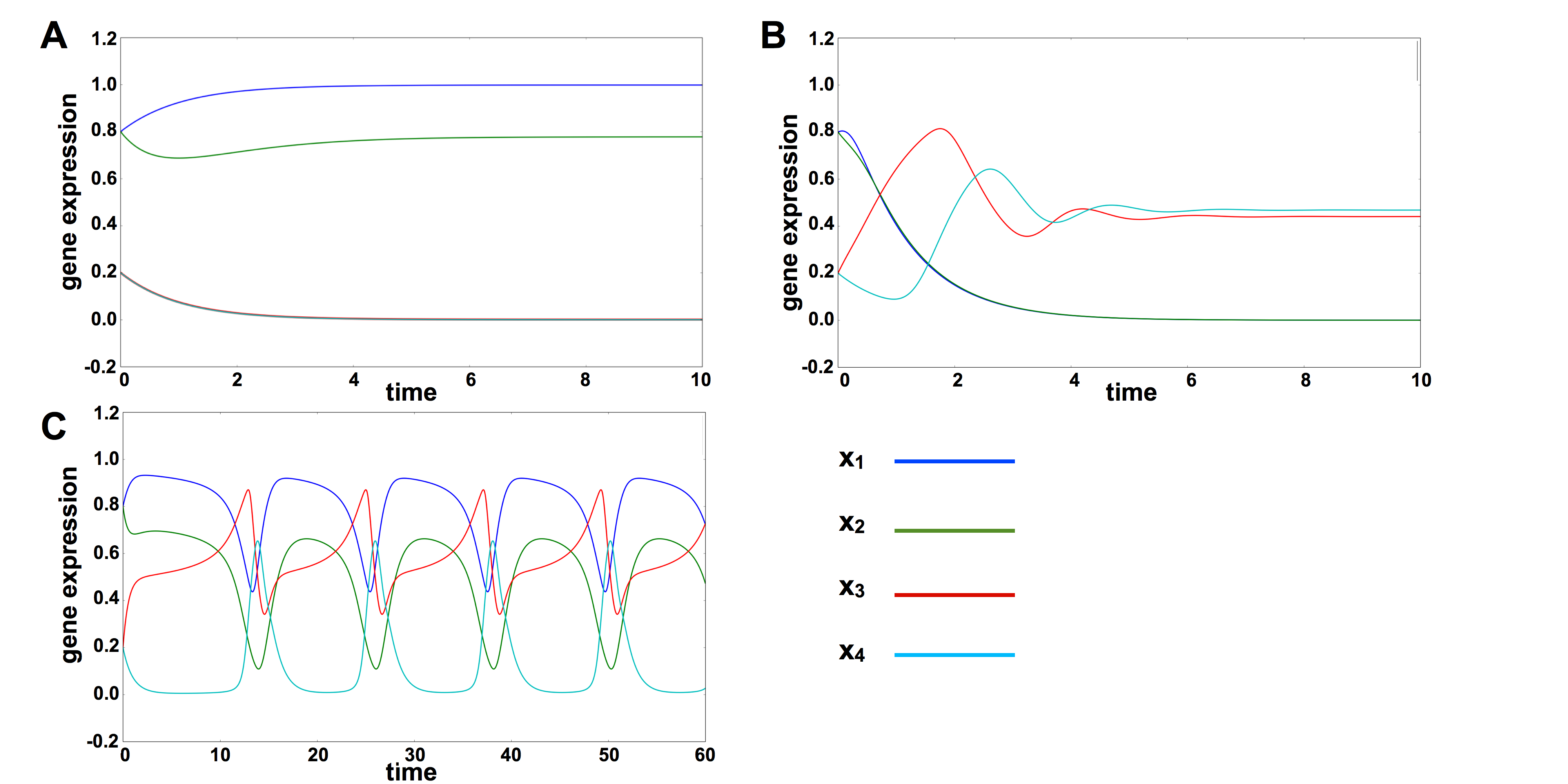}
\caption{{\bf Time series of single-cell dynamics.}
Time series of gene expression for $x_1$, $x_2$, $x_3$, and $x_4$. Each colored line represents expression levels of a single gene. Three different behaviors appeared in single-cell dynamics depending on the parameter $K_{ij}$. We set the parameter $K_{13}$ at (A) $0.98$, (B) $0.58$, and (C) $0.78$. The other parameters were fixed at $K_{34}=0.45, K_{31}=0.94, K_{11}=0.35, K_{21}=0.80$, $K_{42}=0.30$, and $K_{43}=0.45$. A: The pluripotent genes $x_1$ and $x_2$ were highly expressed, and the differentiation genes $x_3$ and $x_4$ were suppressed. This state corresponds to FP. B: Pluripotent genes were suppressed, and differentiation genes were expressed. This state corresponds to FD. C: Oscillation of gene expression occurred, and this state corresponds to O.}
\label{Fig2}
\end{figure}

Using the four-gene model (Fig.~\ref{Fig1}C), we first present the behavior of single-cell dynamics. Depending on the parameter $K_{ij}$, which gives the strength of activation or inhibition from gene $j$ to gene $i$, there are three possible behaviors: (i) fixed-point attractor with high expression of pluripotent genes (FP), given by a fixed-point $x_1 \sim 1$; (ii) fixed-point attractor with high expression of differentiation genes (FD), given by a fixed-point $x_1 \sim 0$; and (iii) the oscillatory state (O), in which all expression levels show temporal cycles (Fig.~\ref{Fig2}). These three states appear as attractors depending on the parameter values $K_{ij}$.

Because the expression level of pluripotent gene $x_1$ is most important for determining the three states, the regulation of gene $x_1$, which is controlled by the parameter $K_{1j}$, is crucial for determining cellular behavior. In particular, threshold $K_{11}$ and $K_{13}$ were found to be critical parameters. Where the value of $K_{11}$ was low, expression of gene $x_1$ was promoted; where the value of $K_{13}$ was low, gene $x_1$ was suppressed. First of all, we set all parameters $K_{ij}$ randomly, and examined the dynamics. If the parameter value of $K_{11}$ ($K_{13}$) was set to a much lower (larger) value (say $0.1$ ($1.0$), respectively), the expression of $x_1$ is fixed to a high value, and the differentiation process would be more difficult. On the other hand, if this parameter value was high (low), $x_1$ was not easily expressed or always expressed, respectively, so that the stem cell state is difficult to be obtained unless other parameter values are finely tuned. With the neighborhood of the above parameters values, the expression level of $x_1$ changes flexibly to other parameters. We then observed that the expression dynamics changed between fixed-point and oscillation easily by changing other parameter values. Indeed, as will be shown, differentiation behavior was observed for a broader range of other parameters. As the parameter space is so huge, we here fixed $K_{11}$ and $K_{13}$ at these values and drew the phase diagram against other parameters. For the parameters $K_{11}$ and $K_{13}$, for example $K_{11}=0.35, K_{13}=0.78$, gene expression levels showed oscillation.

\begin{figure}[h]
\includegraphics*[width=13cm, bb=0 0 2560 903]{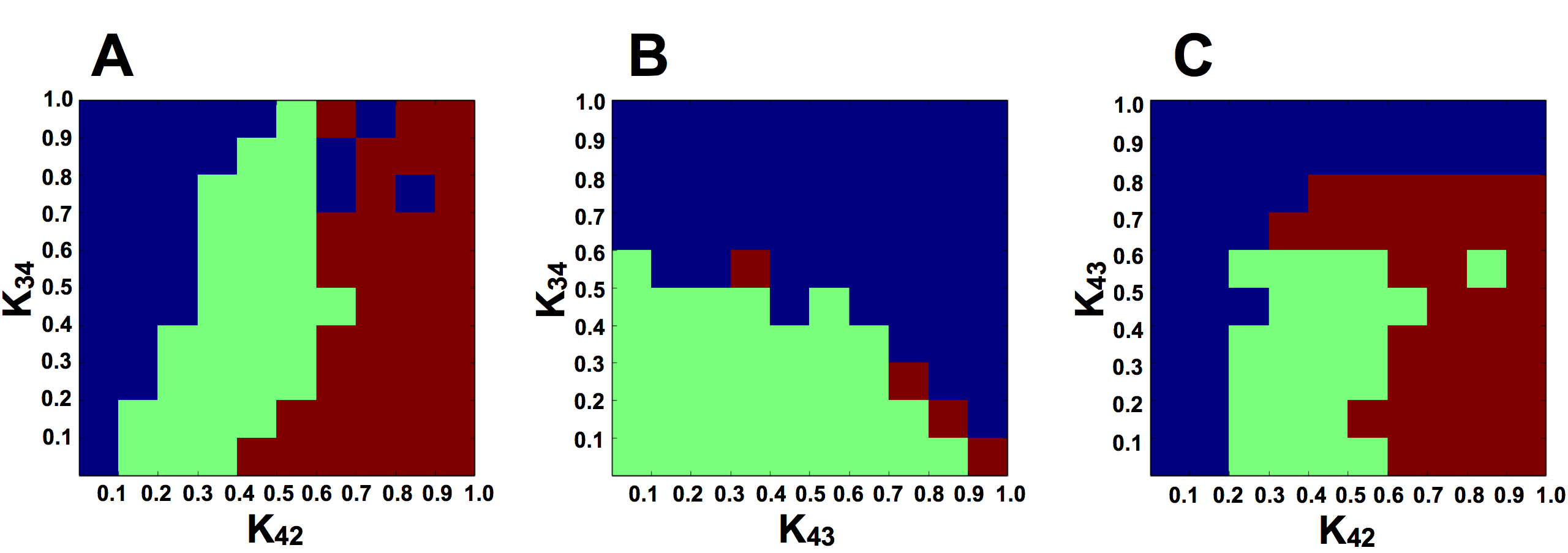}
\caption{{\bf The parameter set and emerging attractors.}
The horizontal and vertical axes represent the values of parameters $K_{34}$, $K_{42}$, and $K_{43}$ (A: $K_{42}$ and $K_{34}$, B: $K_{43}$ and $K_{34}$, C: $K_{42}$ and $K_{43}$). Other parameters were set as in Fig. 2C. Each color represents the type of attractors reached from the initial conditions at pluripotent ($x_1,x_2=0.8$, $x_3,x_4=0$) and differentiated ($x_1,x_2=0$, $x_3,x_4=0.8$) states. Brown, green, and blue represent coexistence of FP and FD, coexistence of O and FD, and existence of FD alone, respectively. For the coexistent cases, the cellular state fell into FP or O by starting from the pluripotent condition, and fell into FD by starting from the differentiated condition.}
\label{Fig3}
\end{figure}

To study FP, FD, and O, i.e., the three states described above, we fixed the parameters $K_{11}$ and $K_{1j}$ (for specific values, see Models), and assessed dependence on the other three parameters $K_{34}$, $K_{42}$, and $K_{43}$ (Fig.~\ref{Fig3}). In most parameter regions, two attractors (stable states) existed, either FP+FD or FD+O depending on the initial conditions. Where the initial condition involved high expression of pluripotent genes, FP or O was reached depending on the parameters; where the initial condition involved high expression of differentiation genes, FD was reached.

For higher values of $K_{34}$ and $K_{43}$, gene-expression oscillation, i.e., the oscillatory state, did not appear, and FP and FD coexisted. Conversely, for lower values of $K_{34}$ and $K_{43}$, the oscillatory state appeared for $0.1 < K_{42} < 0.5$ if pluripotent genes were initially highly expressed. However, where differentiation genes were initially highly expressed, cells fell into FD; thus, FD and O coexisted. As an example of the oscillatory pluripotent state, we fixed the parameters at $K_{34}=0.45$, $K_{42}=0.30$, and $K_{43}=0.45$ for most of the simulations described below.

For oscillatory gene expression, negative feedback is generally required. In our model, negative feedback of gene $x_1$ exists through genes $x_2$, $x_3$, and $x_4$. For the parameter values that generated oscillatory expression, O, auto-promotion and negative feedback of gene $x_1$ (as the key factor in pluripotency) were balanced. Where either of the two regulations was dominant, oscillation ceased and the cellular state fell into either of FP or FD.

\subsection*{Differentiation with cell-cell interaction and noise}
To understand developmental processes, we must investigate the switch from pluripotent to differentiated states. This differentiation event can be mediated either by cell-cell interactions (i.e., by chemicals from other cells) or by noise. Here we explored these two possibilities.

\paragraph{Cell-cell interactions:} Cell-cell interactions play an important role in cellular differentiation. In the simplified GRN we adopted, gene $x_4$ corresponds to \textit{Gata4}. According to Gene ontology database, only \textit{Gata4} among the four genes in the present model concerns with the cell-cell signaling. Hence, we assumed the cell-cell interaction via $x_4$. Indeed, even if other factors $x_i$ ($i \neq 4$) were assumed to diffuse, differentiation by cell-cell interaction to be presented did not appear.

With an increase in the number of cells, differentiation began to occur with specific timing. Following earlier studies\cite{SuzukiN2011,GotoY2013}, we used a model including the cell division process and cell-cell interactions among divided cells. Here, cells divided according to a certain division interval, $t=25$, with the two resultant cells having the same gene expression pattern $x_i$ with the addition of some noise. Starting from a single cell initially in the pluripotent state, i.e., $x_1,x_2 = 0.8$ and $x_3,x_4 = 0.2$, we investigated whether the composition of cells changes. We studied two cases: (A) differentiation from either of FP or FD, and (B) differentiation from the oscillatory state, O.

In (A), where the single-cell state was a fixed point with either expression of pluripotent or differentiated genes, differentiation did not occur by cell-cell interaction, irrespective of the diffusion coefficient $D$. The cellular state remained at the original fixed point.

\begin{figure}[h]
\includegraphics*[width=15cm, bb=0 0 2331 1277]{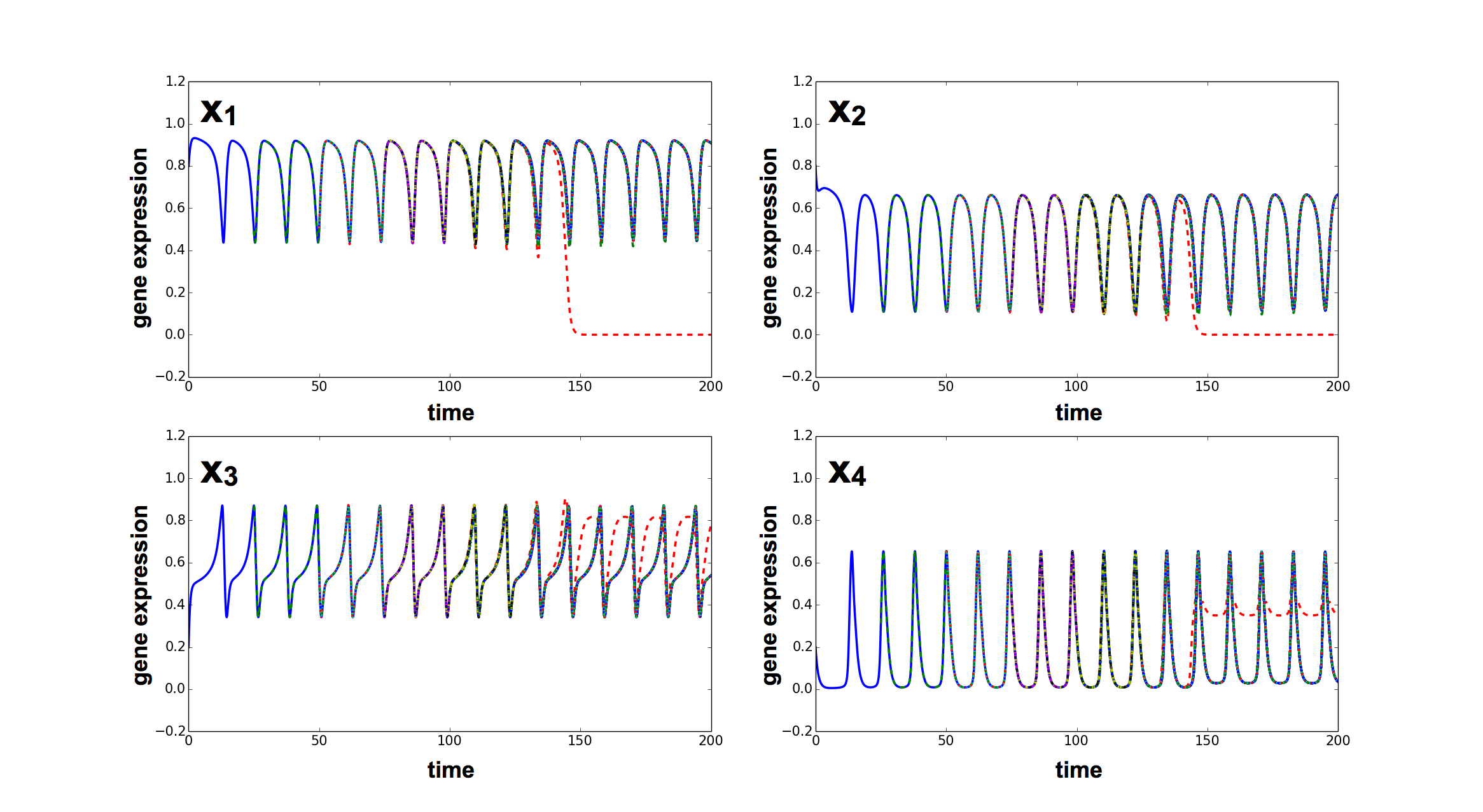}
\caption{{\bf Time series of gene expression with the occurrence of cell-cell interactions.}
Time series of gene expression levels for $x_1$, $x_2$, $x_3$, and $x_4$ for all cells, where cells divided per period $=25$ until time $=125$ to generate $32$ cells. Expression levels of cells are plotted according to color, but most colors are overlaid and, therefore, difficult to discern. The diffusion coefficient $D$ was set at $D=2$, and the other parameter values are the same as those given in Fig. \ref{Fig2}C. The oscillatory state has pluripotency to allow for both self-renewal and differentiation. The oscillation of gene expression was initially desynchronized, and then a few cells switched to the differentiated state.}
\label{Fig4}
\end{figure}

In (B), where differentiation began from the oscillatory gene expression state, with an increase in cell numbers the oscillations of each cell were desynchronized given sufficient strength of cell-cell interactions ($D>2.0$) (Fig.~\ref{Fig4}). With a further increase in the number of cells, some cells lost the oscillation of pluripotent genes, which suppressed the expression of pluripotent genes $x_1$ and $x_2$ and activated expression of differentiation genes $x_3$ and $x_4$. Hence, differentiation and a loss of pluripotency occurred. This process of interaction-induced differentiation from the oscillatory state was first reported by Furusawa and Kaneko \cite{JTB2001}, and then by Suzuki et al. in simpler gene expression network models with fewer genes\cite{SuzukiN2011}. The mechanism of the process can be understood through bifurcation theory\cite{GotoY2013}. After the number of cells reached a given value (e.g., 32), the differentiated and pluripotent cells with oscillatory gene expression coexisted. The ratio of differentiated cells to the total number of cells was almost independent in each run beginning from the oscillatory state, even where noise was included in the division process; thus, proportional regulation of differentiated cell types was achieved. The ratio of differentiated cells increased with the diffusion coefficient of cell-cell interactions $D$, and the time required for differentiation increased with this ratio (\nameref{FigS1}).

\begin{figure}[h]
\includegraphics*[width=13cm, bb=0 0 1949 562]{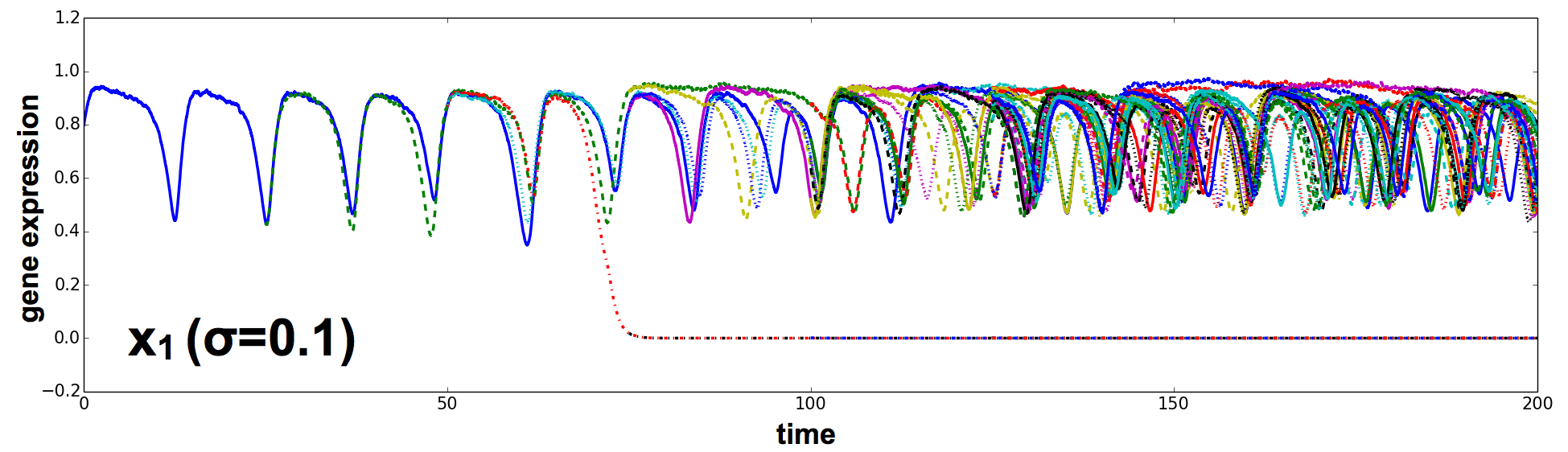}
\caption{{\bf Cellular state transition under noise.}
Time series of gene expression levels for $x_1$. Similar conditions to those described in Fig. \ref{Fig4} were adopted, except that a Gaussian noise term with the amplitude $\sigma=0.1$ was included. Expression levels of cells are plotted according to color. Gene expression oscillation was irregular because of the noise. Irreversible transition from the oscillatory pluripotent to the differentiated state ($x_1 \sim 0$) occurred for $\sigma=0.1$.}
\label{Fig5}
\end{figure}

Stationary cellular states were stable given the inclusion of noise, as long as the noise level was not too high. We also studied the influence of stochastic gene expression by including a Gaussian white noise term $\eta$ with the amplitude $\sigma$ in our model by using Langevin equations (see Models). Here, as long as the strength of noise was not too large ($\sigma = 0.1$), the oscillatory expression dynamics and differentiation ratio were not altered. Even though gene expression dynamics were strongly perturbed and oscillation was not clearly visible, the differentiation process still functioned (Fig.~\ref{Fig5}); hence, differentiation from the pluripotent state (with the oscillatory state) was robust to noise.

\paragraph{Transition by noise:} As an alternative scenario, we considered differentiation as state transition by noise. Here, however, as long as the noise level was not too high ($\sigma < 0.1$), the pluripotent state was stable, and the transition to differentiated state did not occur. For a much higher noise level ($\sigma \sim 0.3$), the pluripotent state was not stable. In this case, the transition to the differentiated state occurred by noise, and this occurred even without cell-cell interactions. However, all cells finally fell into the differentiated state. Thus, pluripotent cells did not remain, and the ratio of pluripotent to differentiated cells was not regulated. In addition, this level of noise might be too high to be considered realistic as a model of gene expression dynamics.

By changing some parameter values in the model, a bifurcation occurred from a fixed point with expression of pluripotent genes to the differentiated state. Therefore, by changing the external condition it is possible to transition from the pluripotent to the differentiated state. However, changing the external condition caused all cells to simultaneously switch to the differentiated state, so that no cells with pluripotency remained. In contrast, given oscillatory gene expression and cell-cell interactions, only some of the cells differentiated from the oscillatory state, while others remained in the pluripotent state. Hence, the coexistence of pluripotent and differentiated cells, with irreversible loss of pluripotency, is possible by starting from the oscillatory state.

\paragraph{Compatibility of oscillatory and stochastic dynamics:} Even if the strength of noise is set at a large value (say, $\sigma=1.0$), the differentiation by cell-cell interaction in our model works well.
Besides the noise during the expression dynamics, we have also added noise in the division process. Indeed, even though the strength of noise in cell division is large (say $\sigma_d=1.0$), the differentiation mechanism in our model still works well (\nameref{FigSextra1}).

\subsection*{Differentiation from the pluripotent state with epigenetic modification}
Cellular differentiation in multi-cellular organisms also involves epigenetic changes, such as histone modification and DNA methylation, which stabilize differentiated states: once differentiated, cells do not regain pluripotency even if the expression level is perturbed. Hence, we introduced epigenetic modification into our model to strengthen the stability of the differentiated state.

Currently, there is no definitive method for introducing the epigenetic process because the precise molecular process of histone modification is difficult to implement in a model with gene expression dynamics. However, it is possible to model the influence of the epigenetic process on expression dynamics phenomenologically\cite{FurusawaKaneko2013, GambarS2014, SasaiM2013, Sneppen2008, Sneppen2007}. The epigenetic process tends to fix the expression state; for example, when a given gene is expressed for a given period, its expression tends to become fixed, and when it is not expressed for a given period, it remains silenced. In other words, the threshold for expression decreases or increases when the gene is expressed or suppressed over a given time span, respectively.

Thus, we introduced epigenetic feedback regulation as a change in the threshold for expression, which was previously given by the expression threshold parameter $K_{ij}$ in our GRN model. Here, we replaced the parameter $K_{ij}$ with an epigenetic variable $\theta_{ij}(t)$, which changes over time depending on expression levels. Consequently, the expression level of the regulator $x_j$ affects that of the regulatee $x_i$ through this epigenetic variable. This is given as dynamics as

\begin{equation}\label{eqepigene}
\dot \theta_{ij}(t) = \frac{1}{\tau_{epi}} (\Theta_{ij} - \theta_{ij}(t) - \alpha x_j(t)).
\end{equation}

The threshold $\theta_{ij}(t)$ is elevated to $\Theta_{ij}$, when the gene $x_j$ is not expressed (i.e., $x_j(t) \sim 0$), whereas the threshold decreases to $\Theta_{ij} - \alpha x_j(t)$ when the gene is fully expressed, i.e., $x_j(t) \sim 1$. Hence, the term $- \alpha x_j(t)$ represents epigenetic feedback, i.e., if gene $x_j$ is expressed, it is more likely to be expressed; if it is not expressed, it is less likely to be expressed. The term $\Theta_{ij}$ thus represents the epigenetic barrier for genes that are not expressed.

The strength of epigenetic fixation given by $\Theta_{ij}$ generally depends on each regulation. Since the expression of pluripotent genes in our model is highly variable, a higher value of $\Theta_{ij}$ is required to fix their expression. Hence, we chose higher $\Theta_{ij}$ values for regulations associated with pluripotent genes. Specifically, epigenetic fixation threshold values were set to $1.0$ for the pluripotent genes $\Theta_{31}$, $\Theta_{21}$, and $\Theta_{42}$, while they were lowered to $0.78$ for the differentiation regulators $\Theta_{13},\Theta_{34},\Theta_{43}$.

For auto-regulation $\Theta_{11}$, the threshold value was set lower, e.g., at $0.50$, since self-activation tightly fixes the expression with small $\Theta_{ij}$. This is because the genes to regulate and to be regulated are identical. This was due to the simplification, which included the self-activation (we examine the five-gene model without self-activation below, in which all $\Theta_{ij}$ for pluripotent genes are set to $1.0$).

Given these parameters, we simulated our model with epigenetic feedback regulation and studied dependence on the epigenetic variables $\tau_{epi}$ and $\alpha$. Initially, we focused on the epigenetic variable $\theta_{ij}(0) = K_{ij}$, which was set with values for cases with (A) fixed-point states and (B) the oscillatory state.

\paragraph{(A) Starting from a state with fixed-points:} In this case, the epigenetic process fixed its state, and no more changes were induced. Thus, in the case of bistable fixed-points, the behavior was almost the same as in the non-epigenetic model. However, by starting from a fixed point with expression of pluripotent genes, the expression began to oscillate via the epigenetic feedback when the time scale $\tau_{epi}$ was small. This did not cause cell differentiation, however, and the oscillation soon ceased before the cell differentiated. Hence, the effect of epigenetic variables is minor in the bistable fixed-point case.

\begin{figure}[h]
\includegraphics*[width=15cm, bb=0 0 2331 1277]{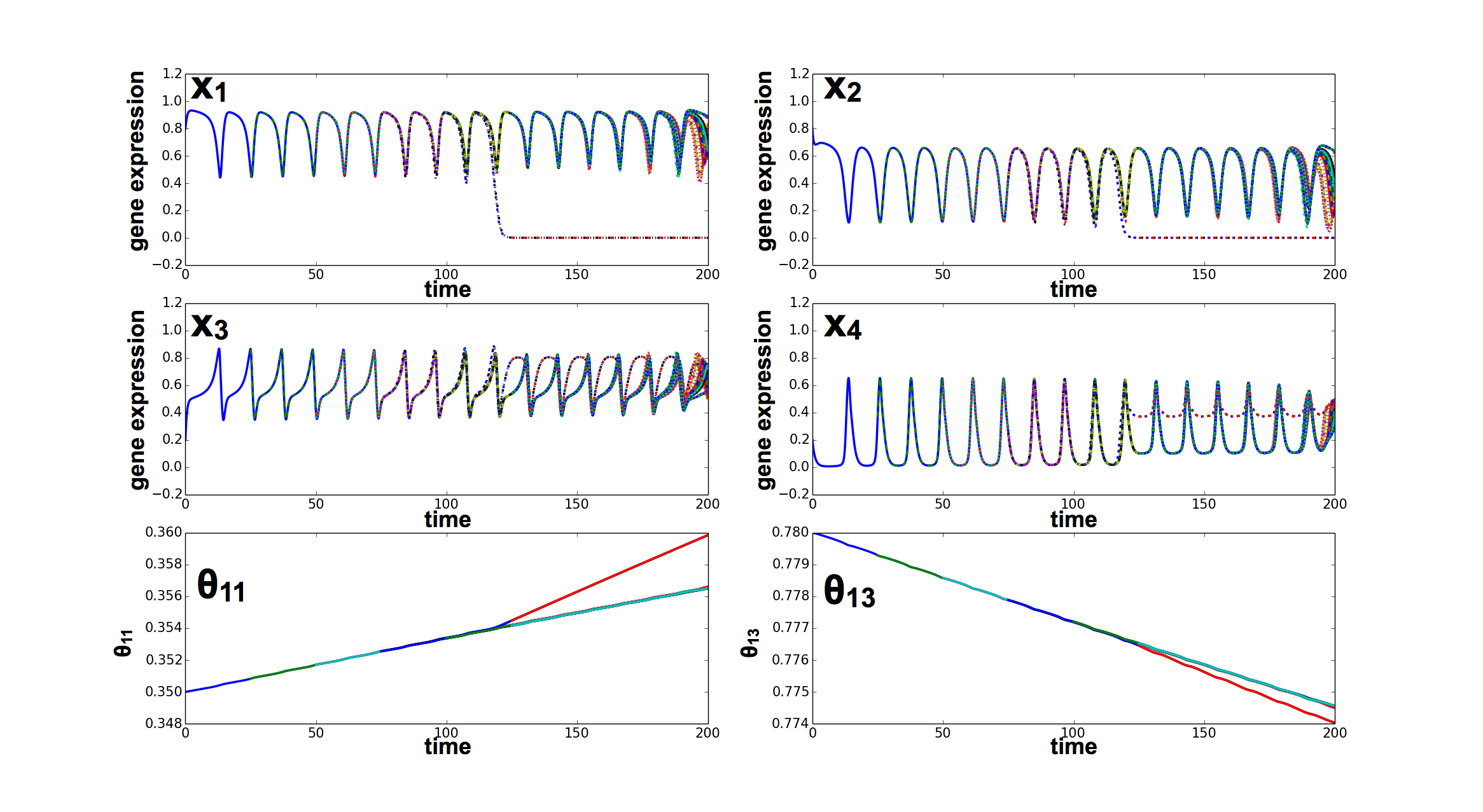}
\caption{{\bf Cell differentiation with the epigenetic variable.}
Time series of gene expression levels for $x_1$, $x_2$, $x_3$, $x_4$, and the epigenetic threshold variables $\theta_{11}(t)$ and $\theta_{13}(t)$. Expression levels of cells are plotted according to color, but most colors are overlaid and, therefore, difficult to discern. We set the parameters of the epigenetic variable as follows: $\tau_{epi}=2.0 \times 10^{3}, \alpha=0.1, \Theta_{ij}=1.0$. The initial value of the epigenetic variable $\theta_{ij}(0)$ was set as $K_{ij}$ in the non-epigenetic model. The other parameters are the same as those used in Fig. 4. First, gene expression oscillated, and then the epigenetic variables in each cell changed gradually. $\theta_{11}(t)$ differentiated into two groups, and in one of these $x_1$ approached $0$.}
\label{Fig6}
\end{figure}

\begin{figure}[h]
\includegraphics*[width=13cm, bb=0 0 3803 1240]{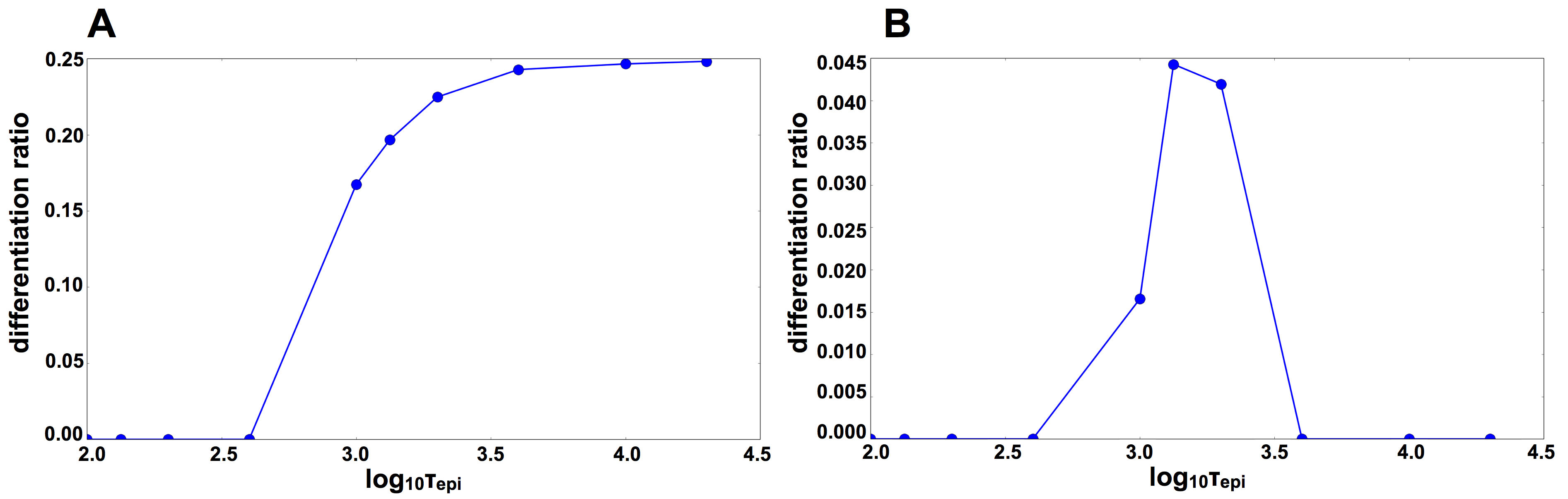}
\caption{{\bf Effect of the time scale of the epigenetic variable $\tau_{epi}$.}
Here, the differentiation ratio is plotted against $\tau_{epi}$. We conducted a differentiation simulation by running the epigenetic model $1000$ times for each time scale, and then counted the number of differentiated cells ($x_1 \sim 0$). The average differentiation ratio (i.e., the vertical axis) represents the percentage of differentiated cells per simulation. To run the simulations, we used identical values to those used in Fig. 6. A: Given strong cell-cell interaction ($D=2.0$), our model showed differentiation without the epigenetic variable. Thus, a large time scale did not negatively affect differentiation. However, given a smaller time scale, cellular differentiation did not occur because the cellular state was quickly fixed. B: Given weak cell-cell interaction ($D=1.5$), there was a peak around the time scale of the epigenetic variable $\tau_{epi} \sim 5.0 \times 10^{3}$. If the time scale of $\tau_{epi}$ was small ($\tau_{epi} < 10^{3}$), the cellular state was quickly fixed and differentiation did not occur. Conversely, if the time scale $\tau_{epi}$ was large ($\tau_{epi} > 10^{4}$), oscillation remained in place.}
\label{Fig7}
\end{figure}

\paragraph{(B) Starting from the oscillatory state:} If the time scale of $\tau_{epi}$ was not too high (the range is discussed below), the differentiation process initially progressed in the same manner as observed without the epigenetic process. Later, however, the cellular state was fixed at an undifferentiated or differentiated state by the change in the epigenetic variable $\theta_{ij}(t)$. After sufficient time, the oscillation of pluripotent genes was lost, and the ability to differentiate was lost after division (Fig.~\ref{Fig6}). Whether or not differentiation and fixation progressed depended on the time scale $\tau_{epi}$ and the coupling constant $\alpha$ (Fig.~\ref{Fig7}, \nameref{FigS2}).

Even with weak cell-cell interactions (e.g., $D=1.5$), where differentiation did not occur without epigenetic feedback regulation, differentiation was sometimes mediated by epigenetic regulation. For example, for $D=1.5$, oscillation occurred over a period sufficient to produce differentiated cells for $10^3 < \tau_{epi} < 10^4$, but oscillation disappeared, and the capacity for differentiation was lost (Fig.~\ref{Fig7}). If the time scale of the epigenetic variable $\tau_{epi}$ decreased ($\tau_{epi} < 10^{3}$) or increased ($\tau_{epi} > 10^{4}$), differentiation did not occur.

As the interaction strength (D) increased, the range of $\tau_{epi}$ that allowed for differentiation also increased (Fig.~\ref{Fig7}). For $D>2.0$, if the epigenetic fixation process was slow ($\tau_{epi}>10^3$), cellular differentiation was fixed to both gene expression $x_i$ and to epigenetic change $\theta_{ij}$ (Fig.~\ref{Fig7}). However, if the time scale of the epigenetic variable decreased ($\tau_{epi} < 10^3$), the cellular state was quickly fixed by epigenetic change, and differentiation never occurred.

Furthermore, even without cell-cell interactions ($D=0$), cells switched from pluripotent to differentiated states via the epigenetic process. However, in this case, oscillation was later lost for all cells. Thus, all cells fell into the FD state together (with a differentiation ratio of $1.0$), and coexistence of pluripotent and differentiated states was not possible.

The addition of noise did not substantially change epigenetic modification. The time scale of the epigenetic variable $\tau_{epi}$ was typically much larger than that of the noise variable $\tau_{noise}$. Thus, the stochastic variation was averaged out through the epigenetic fixation process, and once the epigenetic change had occurred, expression levels stabilized to reduce the effect of noise. Therefore, the epigenetic process was robust to noise (or further increased the robustness of the model to noise).

In summary, we added the epigenetic variable $\theta_{ij}(t)$, to replace the expression threshold $K_{ij}$. The cellular state was fixed by these variables, and its stabilization was enforced. Even if a large amount of noise was added, the cellular state was not destabilized. Upon external perturbation of gene expression patterns, the cell quickly returned to its original state after the change in the epigenetic threshold. Thus, the epigenetic variable produced stabilization of the cellular state and irreversibility of cell differentiation.

\subsection*{Reprogramming to the pluripotent state}
Mature cells can be dedifferentiated into iPSCs by inducing Yamanaka factors\cite{TakahashiYamanaka2006}. Indeed, in dynamical-systems theory, such recovery of pluripotency was predicted as cellular-state transition from a differentiated fixed-point to the pluripotent oscillatory attractor induced by forced-expression of several genes\cite{JTB2001}. Here we discuss the conditions for reprogramming, i.e., switching cellular states by experimental manipulation to regain pluripotency, in our model, also by taking the reversal of epigenetic fixation into account.

First, we investigated reprogramming in the model without the epigenetic process. In this case, differentiated cells were reprogrammed by externally increasing the expression of the pluripotent genes instantaneously, i.e., increasing the value of $x_1$. Instantaneous increase in the expression was sufficient here, since the cellular state is represented only by the expression levels of $x_i$. In order to stabilize the differentiated states against perturbations and sustain irreversibility of cell differentiation, the classic model including only gene expression dynamics is insufficient (see also \cite{NiklasKJ2015} ). By introducing epigenetic feedback regulation with a different time scale, we succeeded in obtaining the result consistent with reprogramming experiments.

In the model with the epigenetic process, however, differentiated cells were not reprogrammed by an instantaneous increase in $x_i$. Following overexpression, cells quickly returned to the differentiated fixed-point. This is because the epigenetic variable, which cannot be altered over a short period, was already increased so that expression of pluripotent genes could not be recovered by instantaneous, or short-term, overexpression. Indeed, we examined the instantaneous overexpression of each gene, as well as a combination of several genes, but reprogramming never occurred.

\begin{figure}[h]
\includegraphics*[width=6cm, bb= 0 0 508 365]{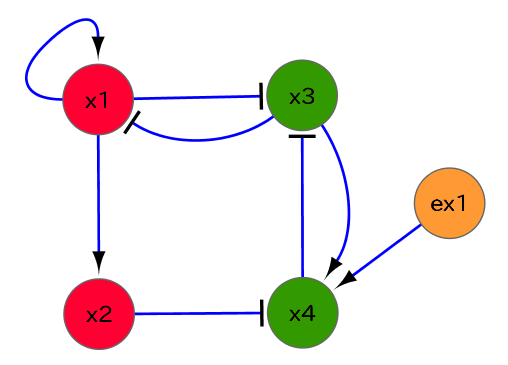}
\caption{{\bf The gene regulatory network in the reprogramming simulation.}
We overexpressed the pluripotent genes $x_1$ and $x_2$, and added an external stimulus $ex1$ to activate gene $x_4$ in the four-gene network model (see Fig.~\ref{Fig1}C). This induction triggered reprogramming, and cells started to oscillate once again. These reprogramming factors correspond with, for example, $x_1$=\textit{Oct4}, $x_2$=\textit{Sox2}, and $ex1$=\textit{Myc}.}
\label{Fig8}
\end{figure}

\begin{figure}[h]
\includegraphics*[width=15cm, bb=0 0 2331 1277]{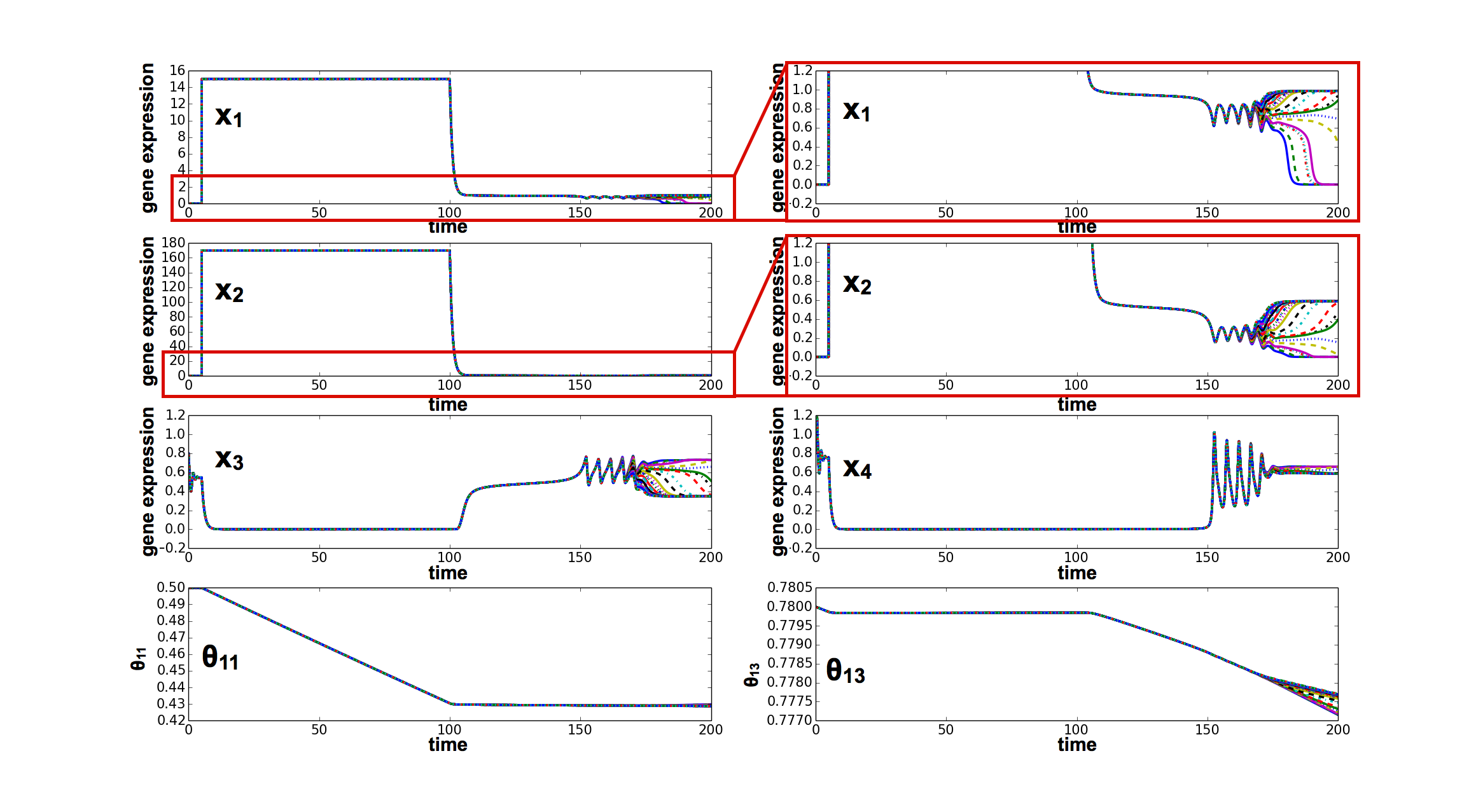}
\caption{{\bf Time series of gene expression in the reprogramming simulation, with induction of pluripotent genes and external activation.}
Plotted here are the time series of gene expression levels for $x_1$, $x_2$, $x_3$, $x_4$, and the epigenetic threshold variables $\theta_{11}(t)$ and $\theta_{13}(t)$. Initially, all cells (e.g., $32$ cells) were set at the differentiated state ($x_{1, 2} = 0, x_{3, 4}=0.8$), with the epigenetic fixation threshold values set at $1.0$ for the pluripotent genes $\Theta_{31}$, $\Theta_{21}$, and $\Theta_{42}$, and at $0.78$ for the differentiation regulators $\Theta_{13},\Theta_{34},\Theta_{43}$. The auto-regulator $\Theta_{11}$ was set at $0.50$. We overexpressed genes $x_1$ and $x_2$ for a long period ($T^e \sim 100$). The epigenetic variables in each cell changed gradually because of the overexpression of these genes. In addition, the gene $x_4$ was promoted by an external stimulus. After overexpression, gene expression began to oscillate again and a few cells showed differentiation. Thus, cells were reprogrammed.}
\label{Fig9}
\end{figure}

We then introduced the overexpression of pluripotent genes into a differentiated cell over a sufficiently long time span $T^e$. For example, pluripotent genes were overexpressed from $t=1$ to $T^e=100$ to the level of $x_i=15$. Additionally, we added external activation of gene $x_4$ to inhibit the expression of gene $x_3$ (Fig.~\ref{Fig8}). In this case, cells were reprogrammed, and gene expression levels regained oscillation and recovered pluripotency. The expression threshold was also reduced (Fig.~\ref{Fig9}), so that epigenetic fixation was relaxed. By starting with this reprogrammed cell, some of the divided cells differentiated given a sufficient level of cell-cell interaction.

\begin{figure}[h]
\includegraphics*[width=15cm, bb=0 0 2331 1277]{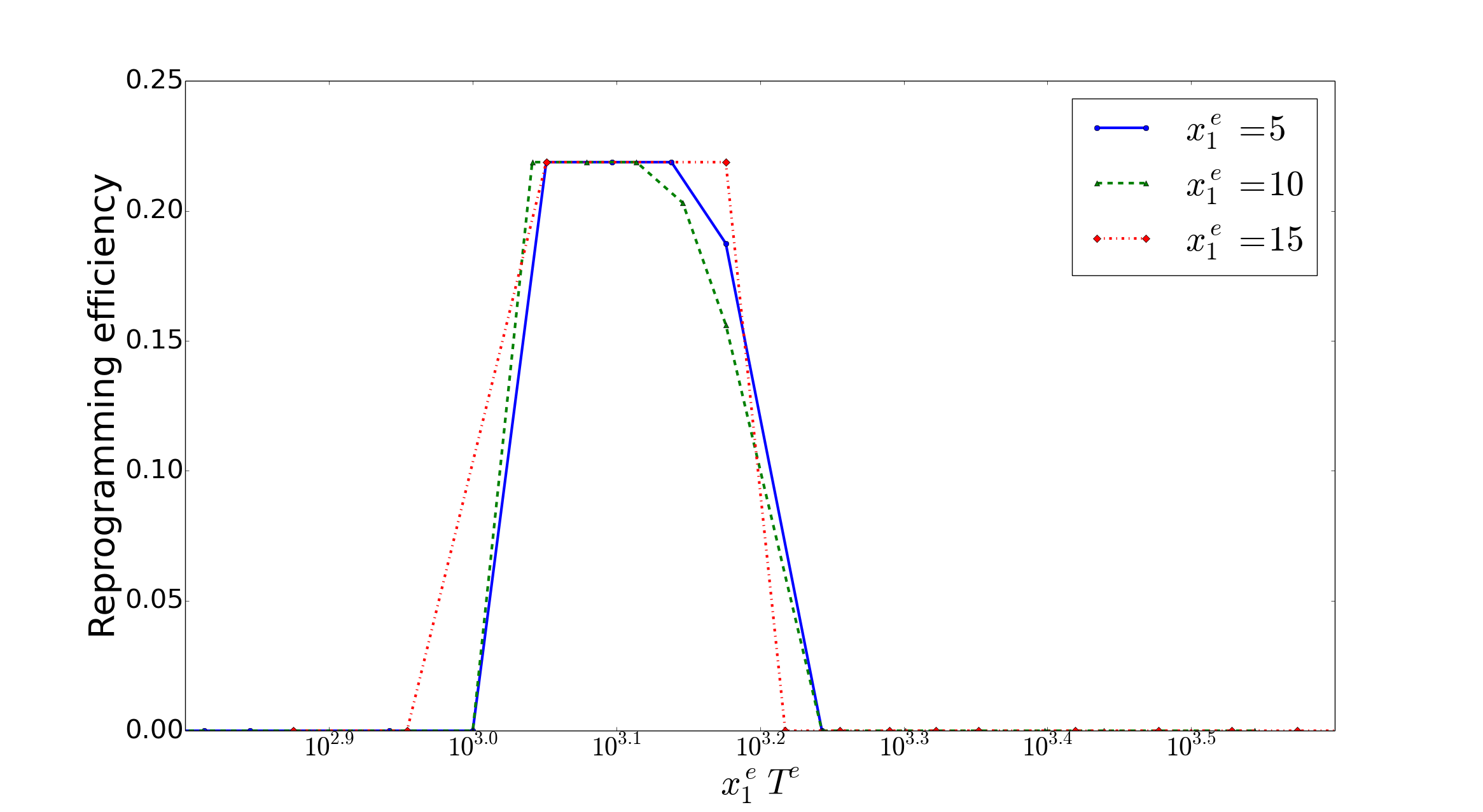}
\caption{{\bf Reprogramming efficiency against overexpression condition.}
Plotted here are Reprogramming efficiency against the product of overexpression value $x^e_1$ for the gene $x_1$ and its duration time $T^e$ semi-log plot. We counted the number of reprogrammed cells (that are differentiated after oscillation). The average reprogramming ratio represents the percentage of reprogrammed cells per simulation. There was a peak around the product $x^e_1 T^e \sim 10^{3.1}$. If the product $x^e_1 T^e$ was small ($x_1^e T^e < 10^3$), cells quickly returned to the differentiated fixed-point. Inversely, if it was large ($x_1^e T^e > 10^{3.3}$), cells fell into the pluripotent fixed-point, and did not show differentiation.}
\label{Figextra2}
\end{figure}

After overexpression of $x_i$ to the value $x^e$ for time span $T^e$, the epigenetic variable $\theta_{ij}(t)$ was estimated to decrease to $\alpha x^e\times \frac{T^e}{\tau_{epi}}$. Hence, epigenetic fixation is relaxed if this value reaches $\theta_{ij}(0)$, where $\theta_{ij}(0)$ is the value after epigenetic fixation. Where $\tau_{epi}=5.0 \times 10^{-4}$ and $\alpha=0.1$, for example, $x^eT^e$ must be larger than $3.0 \times 10^5$ for $\theta_{11}(t)$ to return to the initial value $0.35$. For example, if the overexpression value is changed from $15$ to $3$, overexpression time required about $5$ times. The product of overexpression value and time determines the reprogramming. The reprogramming ratio increases (in a threshold-like manner) as the product increases beyond a critical value $10^3$ (Fig.~\ref{Figextra2}). Indeed, this is natural, as the relaxation process of epigenetic fixation is determined by the product.

The epigenetic fixation is introduced so that the genes that are not expressed are harder to be expressed, following observations in cell-biology. Accordingly, the strength of epigenetic fixation $\Theta_{ij}$ has to be larger than the value of $\theta_{ij}$ chosen initially. Therefore, epigenetic fixation threshold values $\Theta_{ij}$ for pluripotent genes were set to $1.0$ because the maximum value of initial threshold values $\theta_{ij}(0)$ was $0.94$. If it is set to a lower value, the gene is not remained silenced due to the epigenetic change, even when it is not expressed. On the other hand, if the epigenetic fixation threshold values $\Theta_{ij}$ for differentiation regulators were also set to $1.0$, the reprogramming by overexpressing the corresponding genes as well as external factors was not possible. In fact, we carried out both differentiation and reprogramming simulations by choosing a variety of values of $\Theta_{ij}$, and confirmed that epigenetic fixation threshold values for pluripotent genes have to be larger than that for differentiation regulators, to be consistent with experimental observations.

In addition to overexpression levels and time span, the number of overexpressed genes is important. Reprogramming did not occur by overexpression of a single gene, even though its level and time span were sufficient to decrease the epigenetic variable: two or more appropriate genes had to be overexpressed. If a single gene $x_1$ was overexpressed over a sufficient period, the transition to a different fixed-point state occurred, but gene expression did not regain oscillation. By starting from this cell with this new fixed-point state, differentiation did not occur again even when the number of cells was increased. These cells showed increased expression of pluripotent genes up to the level of the original pluripotent cell, but they did not regain the capacity for differentiation. Thus, decreasing the epigenetic threshold variable of pluripotent genes was not sufficient for reprogramming.

We then conducted a reprogramming simulation by changing the initial values for the epigenetic variable $\theta_{ij}(0)$, that is, $\Theta_{ij}$. In general, as $\Theta_{ij}$ became smaller, epigenetic fixation became weaker, and the number of genes that had to be overexpressed decreased. For example, if $\Theta_{34}=0.5$ and $\Theta_{43}=0.3$, the overexpression of just two factors, $x_1$ and $x_2$, without the external overexpression of any other genes could lead to reprogramming (\nameref{FigS3}).

According to our results, pluripotent stem cells had an oscillatory gene expression component; thus, the recovery of oscillation was necessary for recovery of pluripotency. However, oscillation alone was not always sufficient for pluripotency. If the decrease in the epigenetic threshold value was insufficient, the oscillation was weak and the bifurcation to a differentiation fixed point could not occur by cell-cell interactions. In this case, pluripotent genes were expressed. A cellular state such as this, with expression of pluripotent genes but without differentiation potential, corresponds to the pre-iPS state previously reported in reprogramming experiments\cite{MattoutA2011,ChenJ2013}.

\subsection*{The five-gene model}
To promote expression of pluripotent genes, there is an auto-expression loop. This auto-expression is mediated via positive feedback by mutual regulation of genes. In the four-gene model, which has been described and studied thus far, this positive feedback loop was introduced as the self-expression of $x_1$. Auto-expression such as this may be over-simplified, especially considering epigenetic modification as already mentioned. In reality, the auto-expression feedback loop consists of a number of genes. Hence, we replaced auto-regulation of $x_1$ in the four-gene model with a loop structure via a new gene $x_5$ (as shown in Fig.~\ref{Fig1}B), and attempted to validate our previous results and examine the conditions necessary for reprogramming in comparison with experiments.

First, we confirmed that the two fixed-points, FP and FD, and the oscillatory state, O, existed in the five-gene model (see SI, \nameref{TextS1}). Once confirmed, we also included epigenetic threshold variables, as in the four-gene model. For example, we used two epigenetic fixation parameters depending on the regulator type, i.e., the epigenetic fixation value for the pluripotent regulators ($\Theta_{15},\Theta_{31},\Theta_{21},\Theta_{51},\Theta_{42}$) was $1.0$, and for the differentiation regulators ($\Theta_{13},\Theta_{34},\Theta_{43}$) it was $0.65$. Additionally, we confirmed that the switching from oscillatory state to FD progressed via cell-cell interactions (\nameref{FigS5}).

\begin{figure}[h]
\includegraphics*[width=15cm, bb=0 0 2331 1277]{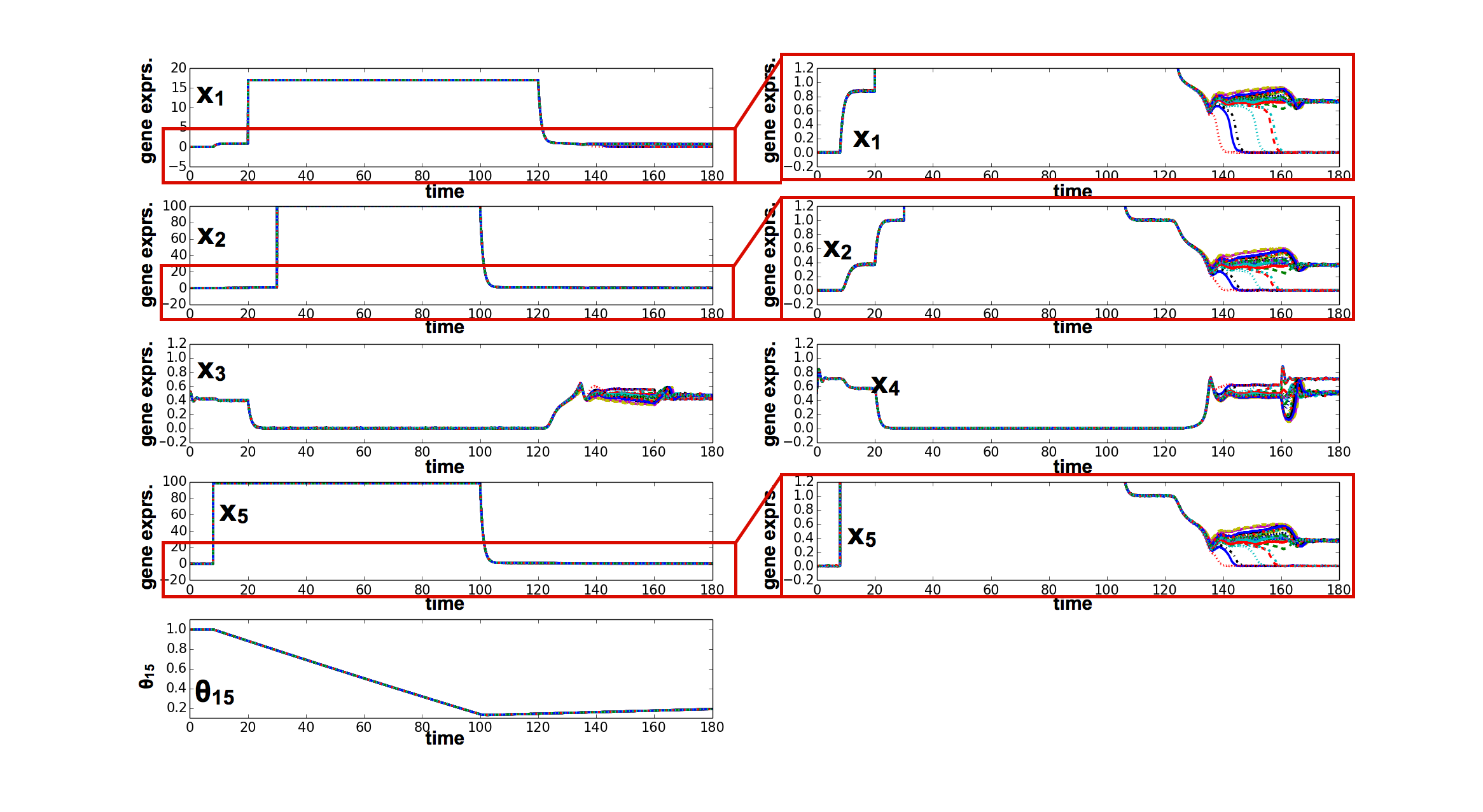}
\caption{{\bf Time series of gene expression in reprogramming via overexpression of three genes and one external factor.}
Plotted here are time series of gene expression for $x_1$, $x_2$, $x_3$, $x_4$, $x_5$, and the epigenetic threshold variables $\theta_{15}$. In this case, we used the following parameter set: $\theta_{13}(0)=\theta_{34}(0)=\theta_{43}(0)=0.65$, $\theta_{15}(0)=\theta_{31}(0)=\theta_{21}(0)=\theta_{51}(0)=\theta_{42}(0)=1.0$. In the differentiated state, we overexpressed genes $x_1, x_2$, and $x_5$ over a long period and added one external regulator. Induction of these factors changed the epigenetic threshold variables; gene expression then began to oscillate again and, later, differentiation occurred in a few cells.}
\label{Fig10}
\end{figure}

To regain pluripotency from the differentiated state, in our reprogramming experiment with the five-gene model, overexpression of the genes $x_1$, $x_2$, and $x_5$, as well as one external factor (to inhibit gene $x_4$), was necessary. These four genes correspond to the Yamanaka factors (\textit{Oct4, Sox2, Klf4}, and \textit{Myc}) used for reprogramming (Fig.~\ref{Fig10}, \nameref{FigS6}). As long as we started the reprogramming simulation after the threshold value $\theta_{ij}(t)$ for differentiated cells reached the pre-set level $\Theta_{ij}$, these four genes were necessary for reprogramming.

The number of genes that had to be overexpressed depended on the level of epigenetic fixation. In general, overexpression of the four aforementioned genes over a sufficient period was required for reprogramming to reset the value of epigenetic variables for the differentiated cells (epigenetic fixation was complete to have $\theta_{ij}(t) \sim \Theta_{ij}$). In contrast, reprogramming was easier if epigenetic fixation was insufficient, and fewer genes, including $x_5$, were sufficient for reprogramming.

\section*{Discussion}
In this study, we assessed a simplified model that was a part of an inferred GRN previously reported by Dunn et al. \cite{SmithAG2014}. Some regulations were simplified by deleting mediator genes, but the core network that is believed to be important for pluripotency, in particular the network motif for a toggle switch, was included. In accordance with the reported GRN, the genes in the model corresponded to \textit{Nanog, Oct4, Gata6}, and \textit{Gata4}, while the additional gene in our five-gene model corresponded to \textit{Klf4}.

We showed that oscillation and switching between high and low levels of gene expression causes some cells to fall into differentiated states via cell-cell interactions. This interaction-induced differentiation from the oscillatory state was robust to noise. Indeed, expression levels of the pluripotency-related gene \textit{Hes1} are reported to oscillate in stem cells, but oscillation is apparently lost after differentiation\cite{Kageyama2013}. This observation is consistent with our oscillation-based mechanism.

Alternative proposals for the differentiation mechanism are based solely on multistability and stochasticity. According to these views, both the differentiated and pluripotent states are given by one of the multi-stable fixed-points, and cellular-state transition is caused by noise. For example, a GRN with auto-promotion and mutual inhibition between two genes\cite{HuangS2011} can produce such bistability. The noise level is critical to this differentiation process. Unless noise level is optimally tuned, the transition between the pluripotent and differentiated states continued to occur via noise, and irreversible differentiation did not occur. Additionally, because switching is stochastic, this model could not control the ratio of pluripotent to differentiated cells, and once a cell was in one of the bistable states, the epigenetic process fixed this state.

In contrast, differentiation from oscillatory dynamics and cell-cell interactions is robust to noise. This provides an explanation for pluripotency as oscillatory dynamics, and characterizes the irreversible differentiation as a transition from oscillatory to fixed-point dynamics, which, later, is consolidated by epigenetic feedback.

In contrast to our findings, however, a recent study suggested that gene expression in stem cells shows stochastic switching between high and low levels, rather than oscillation dynamics\cite{Elowitz2014}. We note that our mechanism works even with strong stochasticity. Even though the strength of noise is set at a large value (say, $\sigma=1.0$), the differentiation by cell-cell interaction in our model works well. Besides the noise during the expression dynamics, we have also studied the noise in the division process. Indeed, even though the strength of noise in cell division is large (say $\sigma_d=1.0$), the differentiation mechanism in our model still works well. Where this is the case, the oscillatory component underlies gene expression that shows noisy dynamics. Hence, the experimental observation did not contradict our oscillation scenario. Under such noise level, the differentiation ratio from sibling is not necessarily correlated as in the experimental results. Under these high noise levels for $\sigma$ and $\sigma_d$, and by setting the parameter values say at $\tau_{div}=12.5$ and $D=1.5$, about $4$ switching occurred per $100$ cell division, as is consistent with the experimental data, while preserving the stochastic oscillatory dynamics.

To check the possibility of stochastic oscillation experimentally, one would need to examine whether an oscillatory component exists among noisy dynamics. This would be possible by measuring the transition probability among three states (A, B, C) and examining if the probability $P(A \rightarrow B)$ has a circulation component, as characterized by the deviation between $P(A \rightarrow B) P(B \rightarrow C) P(C \rightarrow A)$ and $P(B \rightarrow A) P(A \rightarrow C) P(C \rightarrow B)$. We also suggest that by measuring expression of pluripotent genes for a number of iPS cells by single-cell-PCR, one could uncover the loci of oscillatory attractor, as the phase of oscillation is expected to be scattered by cells.

Second, in the experiment of \cite{Elowitz2014} switching between Nanog-high to Nanog-low is less frequent than the result presented here. However, this switching frequency can be easily changed in our model by changing the parameter values $\tau_{div}$, the strength of cell-cell interaction $D$ and noise $\sigma$.

Here, we also introduced epigenetic threshold variables to fix differentiated cellular states via epigenetic changes. The epigenetic variables in our study promoted gene expression if the regulator gene was highly expressed. Conversely, they inhibited gene expression if the expression of the regulator was low. Indeed, epigenetic modification represented by histone modification is known to reinforce gene function by reconstruction of chromatin\cite{Gaspar-MaiaA2011}. For example, the maintenance of pluripotency is promoted and suppressed by open and closed chromatin states in cell differentiation, respectively. The epigenetic feedback process in our model was a mathematical representation of such reinforcement.

In our model, the time scale of epigenetic change $\tau_{epi}$ was much slower than the time scale of gene expression dynamics, by a factor of $10^2-10^3$. Therefore, because the time scale for transcription is seconds to minutes, epigenetic modification appears to occur over days. If the time scale for cell division is hours, the time scales for gene regulation $\tau_{gene}$, cell division $\tau_{div}$, and epigenetic variable $\tau_{epi}$ satisfy $\tau_{gene}<\tau_{div}<\tau_{epi}$. Indeed, in our model, epigenetic fixation of cell differentiation works effectively given these conditions.

If differentiation occurs, and the differentiation ratio depends on the time scale of epigenetic modification, the rate of epigenetic change can control the distribution of cell types. Hence, epigenetic fixation controls cell distribution and is, therefore, essential to the stabilization of cellular states.

However, epigenetic fixation also provides a barrier in reprogramming. In contrast to the scenario without epigenetic fixation, simply resetting gene expression patterns is not sufficient to reprogram differentiated cells. Even if the gene expression pattern of a differentiated cell is reset to the pluripotent state, the cellular state quickly returns to a differentiated state because of the change in the epigenetic threshold variables.

Reprogramming also requires overexpression of pluripotent genes over a time span of $\tau_{epi}$. Even with overexpression of the correct genes, an insufficient amount of time cannot relax the epigenetic threshold, and cells quickly return to the differentiated state. Indeed, in reprogramming experiments, Yamanaka factors are overexpressed for days by using retroviruses, during which time, it is suggested that chromatin is reconstructed.

In our model, the overexpression of multiple transcription factors, including pluripotent genes, was generally necessary for reprogramming to occur. Indeed, in the five-gene model, the four factors required for reprogramming were the Yamanaka factors, \textit{Oct4, Sox2, Klf4}, and \textit{Myc}, which are adopted in iPS construction. Even though the GRN in our model contained only five genes, reprogramming required these four factors. In particular, \textit{Klf4} was a prerequisite for reprogramming. In iPS construction, \textit{Klf4} also plays an important role in promotion of reprogramming by interacting with \textit{Oct4} and \textit{Sox2}\cite{WeiZ2009}.

Note that the reprogramming efficiency in experiments is rather low. This might be related with a limited range in the overexpression level in Fig.~\ref{Figextra2}. However, at the moment, it is uncertain if this low efficiency is due to difficulty in adjusting such range of overexpression levels, or due to underlying noisy dynamics, or due to some other experimental constraint.

Experimentally, reprogramming is reportedly easier if the epigenetic fixation of some genes is weaker. Indeed, epigenetic fixation levels depend on the derived cell type or chromatin structure\cite{ErnstJ2011,MarstrandTT2014}. Furthermore, highly efficient reprogramming, such as deterministic (or non-stochastic) reprogramming from the privileged somatic cell state\cite{RaisY2013,GuoS2014}, includes a chromatin remodeling factor or specific types of derived cell. This scheme is expected to relax the level of epigenetic fixation for some genes. Thus, it is consistent with the ease of reprogramming caused by reducing epigenetic fixation parameters $\Theta_{ij}$ for some genes ($j$) in our model.

Our study also demonstrates that cells fall into a fixed state with the expression of pluripotent genes when there is insufficient overexpression to suppress differentiation genes. Gene expression levels in such cells do not show oscillation, nor do cells show differentiation again. Even though pluripotent genes are expressed, the potential for differentiation is not regained. These cells are regarded as being in a pre-iPS state, which was previously reported in reprogramming experiments\cite{MattoutA2011,ChenJ2013}. In these experiments, following overexpression of the Yamanaka factors, the cell did not regain pluripotency even though ES cells-markers (\textit{SSEA-1} and \textit{Oct4}) were expressed.

The GRNs we studied here are based on several experimental reports. In reality, the GRNs responsible for pluripotency and differentiation involve many more components, and other candidate GRNs have also been proposed for pluripotency\cite{BoyerLA2005}. Our conclusions with regard to oscillation-based differentiation, epigenetic fixation, and reprogramming, however, remain valid as long as the present core network is preserved. Additionally, in a differentiation process including the core network structure consisting of \textit{Nanog, Oct4, Gata6}, and \textit{Gata4}, as discussed here, the four factors are required for reprogramming, independently of the parameter values, as is consistent with experimental observations.

In summary, in our study, oscillatory gene expression produced the pluripotency of cells, and differentiation occurred via a state transition to a fixed-point with the suppression of pluripotent genes. These expression patterns were then fixed epigenetically. In our model, differentiation and reprogramming were interpreted as creation (deletion) of gene expression oscillation and the enhancement or relaxation of epigenetic fixation, respectively. Pluripotent states involved the oscillation of expression of several (here, four to five) genes, while differentiated states suppressed the expression of these genes to reduce oscillation. Thus, our results showed that reprogramming to recover pluripotency involves recovery of gene expression, achieved by overexpression of several genes, and relaxation of epigenetic fixation.

\section*{Models}
\subsection*{Construction of GRN model}
The simplified models consisted of either four or five genes with seven or eight regulations, respectively. In simplification of GRN, we decreased nodes and edges as long as the differentiation is possible. As a result, we extracted a four-factor model, as a minimal structure showing differentiation. Furthermore, the reprogramming simulation from this network, as presented in the present paper, is also consisted with experiments. The existence of these regulations in the constructed GRN is supported by earlier studies\cite{BoyerLA2005,CharronF1999,ChazaudC2006,PikkarainenS2004,LiC2013}. In the four-gene model, the self-activating gene (promotion-loop structure) and its cofactor were regarded as pluripotent genes, and the genes inhibited by these pluripotent genes were regarded as differentiation genes. For example, genes $x_1$, $x_2$, $x_3$, and $x_4$ corresponded with \textit{Nanog, Oct4, Gata6}, and \textit{Gata4}, respectively. In the five-gene model, the additional gene was \textit{Klf4}. Among these genes, only \textit{Gata4} functions in cell-cell signaling (interaction) according to Gene Ontology. Hence, we considered cell-cell interactions through diffusive coupling by the gene product of $x_4$.

We introduced epigenetic feedback regulation into the model as a change in the threshold for gene expression. This depends on the expression levels of a regulator gene. If the regulator gene is highly expressed, expression of the regulated gene is promoted; however, where expression of the regulator gene is low, expression of the regulated gene is inhibited. Epigenetic change occurs via the change in threshold for expression dynamics and, with feedback, the cellular state is fixed.

\subsection*{Four-gene model}
Here, we describe our gene expression dynamics model. Cellular states are represented by the gene expression pattern of four genes, $x_1$, $x_2$, $x_3$, and $x_4$. These genes regulate the expression levels of themselves and other genes. Additionally, we consider the expression dynamics of gene $i$ of the $k$th cell at time $t$, denoted as $x_i^k(t)$. Only gene $x_4$ is involved in a cell-cell interaction, which is the diffusion of the gene expression level of $x_4$. Hence, our differential equation model is as follows:
\begin{displaymath}
\left\{
\begin{array}{l}
\dot x_1^k(t) = \frac{(\frac{x_1^k(t)}{K_{11}})^n}{1+(\frac{x_1^k(t)}{K_{11}})^n}\frac{1}{1+(\frac{x_3^k(t)}{K_{13}})^n} - x_1^k(t) \left(+ \eta_1^k\sqrt{x_1^k(t)}\right)\\ \\
\dot x_2^k(t) = \frac{(\frac{x_1^k(t)}{K_{21}})^n}{1+(\frac{x_1^k(t)}{K_{21}})^n} - x_2^k(t) \left(+ \eta_2^k\sqrt{x_2^k(t)}\right)\\ \\
\dot x_3^k(t) = \frac{1}{1+(\frac{x_1^k(t)}{K_{31}})^n}\frac{1}{1+(\frac{x_4^k(t)}{K_{34}})^n} - x_3^k(t) \left(+ \eta_3^k\sqrt{x_3^k(t)}\right)\\ \\
\dot x_4^k(t) = \frac{1}{1+(\frac{x_2^k(t)}{K_{42}})^n}\frac{(\frac{x_3^k(t)}{K_{43}})^n}{1+(\frac{x_3^k(t)}{K_{43}})^n} - x_4^k(t) + \frac{D}{N(t)}\sum_j (x_4^j(t)-x_4^k(t)) \left(+ \eta_4^k\sqrt{x_4^k(t)}\right)\\ \\
\end{array}
\right.
\end{displaymath}
where $D$ is the diffusion coefficient, $\eta_i^k$ is an uncorrelated Gaussian white noise term when the stochastic experiment is considered, and $N(t)$ is the total number of cells at time $t$. Depending on the parameter $K_{ij}$, which gives the strength of activation or inhibition from gene $j$ to $i$, the behavior of our model changes.

In cell division, two new cells are produced that have the same gene expression pattern as the original cell. Additionally, gene expression is slightly perturbed by adding 
a Gaussian white noise $(\sigma_d =1.0 \times 10^{-3})$ after cell division, as  $\eta_i x_i$ or $(1 - \eta_i) x_i$ (with $\eta$ as a random number in $\left[ 0, \sigma_d \right]$ after each cell division.

\subsection*{Five-gene model}
In the four-gene model, the positive feedback loop of the pluripotent gene $x_1$ is introduced for self-activation. Auto-regulation such as this may be over-simplified; in reality, this should be replaced by a feedback regulation loop including a number of genes. Therefore, we change the auto-expression of gene $x_1$ in the four-gene model to a loop structure via the gene $x_5$, and the five-gene model is described as follows:
\begin{displaymath}
\left\{
\begin{array}{l}
\dot x_1^k(t) = \frac{(\frac{x_5^k(t)}{K_{15}})^n}{1+(\frac{x_5^k(t)}{K_{15}})^n}\frac{1}{1+(\frac{x_3^k(t)}{K_{13}})^n} - x_1^k(t) \left(+ \eta_1^k\sqrt{x_1^k(t)}\right)\\ \\
\dot x_2^k(t) = \frac{(\frac{x_1^k(t)}{K_{21}})^n}{1+(\frac{x_1^k(t)}{K_{21}})^n} - x_2^k(t) \left(+ \eta_2^k\sqrt{x_2^k(t)}\right)\\ \\
\dot x_3^k(t) = \frac{1}{1+(\frac{x_1^k(t)}{K_{31}})^n}\frac{1}{1+(\frac{x_4^k(t)}{K_{34}})^n} - x_3^k(t) \left(+ \eta_3^k\sqrt{x_3^k(t)}\right)\\ \\
\dot x_4^k(t) = \frac{1}{1+(\frac{x_2^k(t)}{K_{42}})^n}\frac{(\frac{x_3^k(t)}{K_{43}})^n}{1+(\frac{x_3^k(t)}{K_{43}})^n} - x_4^k(t) + \frac{D}{N(t)}\sum_j (x_4^j(t)-x_4^k(t)) \left(+ \eta_4^k\sqrt{x_4^k(t)}\right)\\ \\
\dot x_5^k(t) = \frac{(\frac{x_1^k(t)}{K_{51}})^n}{1+(\frac{x_1^k(t)}{K_{51}})^n} - x_5^k(t) \left(+\eta_5^k\sqrt{x_5^k(t)}\right) \\
\end{array}
\right.
\end{displaymath}

\subsection*{Epigenetic feedback regulation}
In the equation for $x_i^k(t)$, each parameter $K_{ij}$ is replaced by the epigenetic variable $\theta_{ij}(t)$, which changes over time depending on gene expression levels by introducing epigenetic feedback regulation as a change in the threshold for gene expression as follows:
\[ \dot \theta_{ij}(t) = \frac{1}{\tau_{epi}} (\Theta_{ij} - \theta_{ij}(t) - \alpha x_j(t)), \]
where $\Theta_{ij}$ is the threshold value after epigenetic fixation and $\tau_{epi}$ is the time scale of the epigenetic variable. The value of the epigenetic variable $\theta_{ij}(t)$ changes depending on the expression levels of the regulator gene $x_j$; if the regulator gene $x_j$ is highly expressed, expression of the regulatee gene $x_i$ is promoted, but if expression of the regulator $x_j$ is low, the regulatee $x_i$ is inhibited.

\subsection*{Model parameters}
To numerically investigate our model, we set the Hill coefficient as $n=6$ and $n=4$ in the in four-gene and five-gene models, respectively. The time of cell division $t_{div}$ was chosen to be 25. The results of the model do not depend on these parameters, as long as the former is sufficiently large (e.g., $n \ge 6$) and latter not too large (e.g., $t_{div}<1000$). The maximum number of cell divisions is $5$; hence, the maximum number of cells is $32$.

For most simulations, we used the parameters $K_{ij}$ as follows: $K_{13}=0.78, K_{34}=0.45, K_{31}=0.94, K_{11}=0.35, K_{21}=0.80, K_{42}=0.30$, and $K_{43}=0.45$. In the five-gene model, the additional regulations were $K_{15}=0.14$ and $K_{51}=0.80$. The parameters for epigenetic feedback regulation $\tau$ and $\alpha$ were set to $2.0 \times 10^3$ and $0.1$, respectively.

\subsection*{Model simulation}
By using the code written by C/Python simulations were carried out by using standard Runge-Kutta algorithm.

\section*{Acknowledgments}
The authors would like to thank Tetsuhiro Hatakeyama, Benjamin Pfeuty, and Nen Saito for useful discussions.

\nolinenumbers

\section*{Supporting Information}

\subsection*{S1 Text}
\label{TextS1}
{\bf Supplementary information about the five-gene model.}

Similar to the four-gene model, depending on the dominance between the positive and negative feedback of gene $x_1$, the five-gene model showed three behaviors: (i) fixed-point attractor with high expression of pluripotent genes (FP), (ii) fixed-point attractor with high expression of differentiation genes (FD), and (iii) the oscillatory state (O). The five-gene model also showed differentiation from the oscillatory state (Fig. S4). The attractor depended on the parameters $K_{ij}$ for each edge, while most effective regulations to determine the type of attractors were related to gene $x_1$, as in the four-gene model. In the five-gene model, the expression level of gene $x_5$ was important because the key gene $x_1$ is promoted via $x_5$.

Where oscillatory gene expression was observed, the gene expression levels in each cell were desynchronized with cell division, and the epigenetic threshold variables took different values according to cells. With the change in epigenetic variables, oscillation was attenuated or terminated for some cells, leading to differentiation.

\subsection*{S1 Fig}
\label{FigS1}
\includegraphics*[width=13cm, bb=0 0 3877 1220]{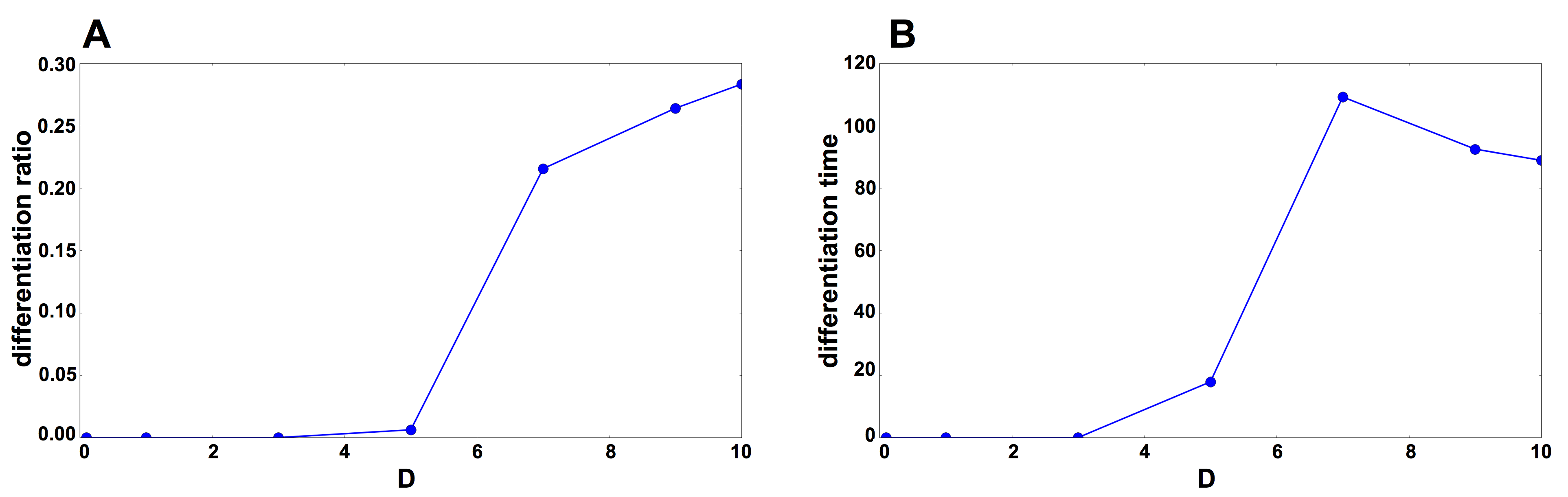}\\
{\bf S1 Fig. Effects of the value of the diffusion coefficient $D$.} The differentiation ratio and time is plotted against the diffusion coefficient $D$. The simulation in the four-gene model was conducted with cell-cell interactions, and with an increase in cell numbers. From $1000$ samples, we counted the number of differentiated cells ($x_1 \sim 0$) for each diffusion coefficient $D$. A: The average of the differentiation ratio (vertical axis) was computed as the percentage of differentiated cells per simulation. Starting from a pluripotent state, the number of cells that went to a fixed point $x_1 \sim 0$ increased with the diffusion coefficient $D$. B: The average time needed for cells to differentiate was computed and plotted as a function of $D$. The time to differentiation increased with the diffusion coefficient $D$.

\subsection*{S2 Fig}
\label{FigSextra1}
\includegraphics*[width=13cm, bb=0 0 1942 574]{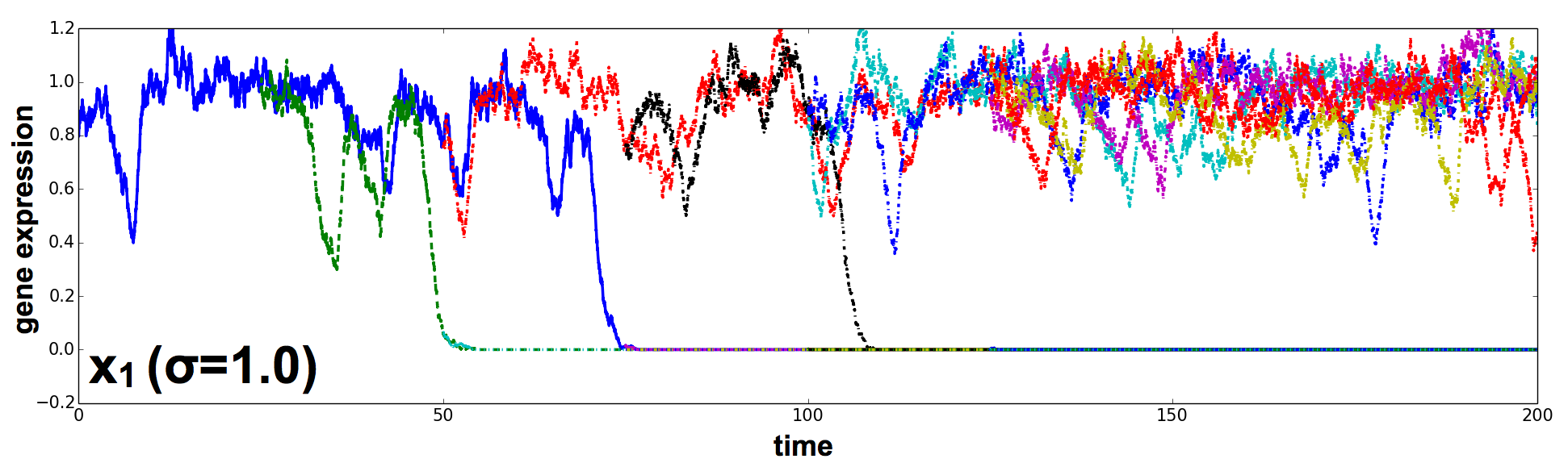}\\
{\bf S2 Fig. Cellular state transition under highly noise.} Time series of gene expression levels for $x_1$ (as in Fig. 5). Similar conditions to those described in Fig. 4 were adopted, except that a Gaussian noise term with the amplitude $\sigma=1.0$ was included. Expression levels of cells are plotted according to color. Gene expression oscillation was irregular because of the noise. Irreversible transition from the oscillatory pluripotent to the differentiated state ($x_1 \sim 0$) occurred for $\sigma=0.1$.

\subsection*{S3 Fig}
\label{FigS2}
\includegraphics*[width=11cm, bb=0 0 1721 1240]{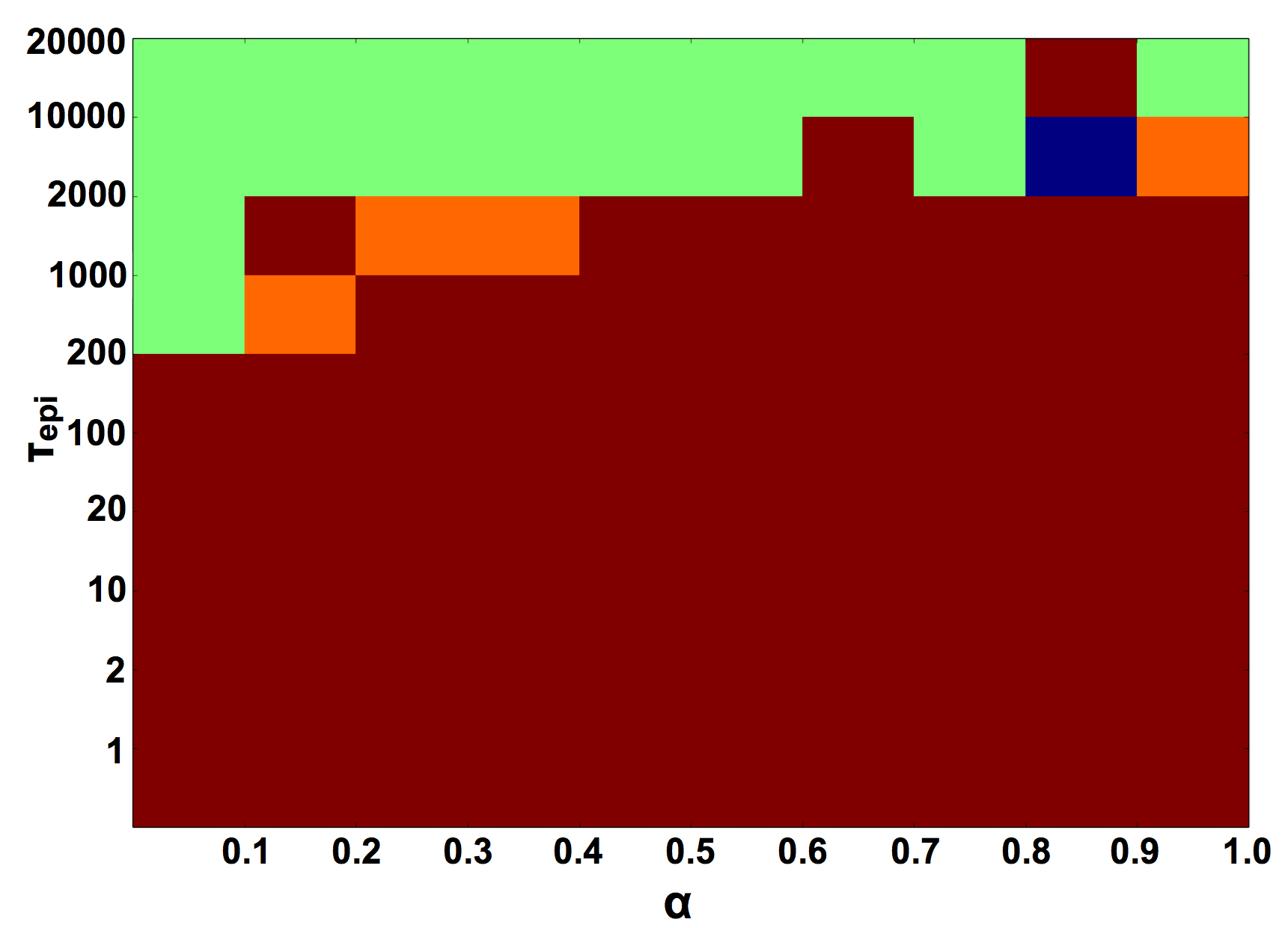}\\
{\bf S3 Fig. Phase diagram of differentiation behavior plotted against time scale $\tau_{epi}$ (vertical axis) and $\alpha$ (horizontal axis).} Parameters were set as in Fig. 6. Brown, orange, green, and blue indicate fixed points at $x_1=0$ without differentiation, differentiation and loss of oscillation, preservation of oscillation, and the fixed-point of the differentiated state ($x_1 \sim 0$) (FD), respectively. For low values of $\tau_{epi}$, the epigenetic fixation progressed quickly, and then cells reached the fixed-point with expressed pluripotent genes (FP). Differentiation from the oscillatory state appeared at a high value of $\tau_{epi}$.

\subsection*{S4 Fig}
\label{FigS3}
\includegraphics*[width=15cm, bb=0 0 2331 1277]{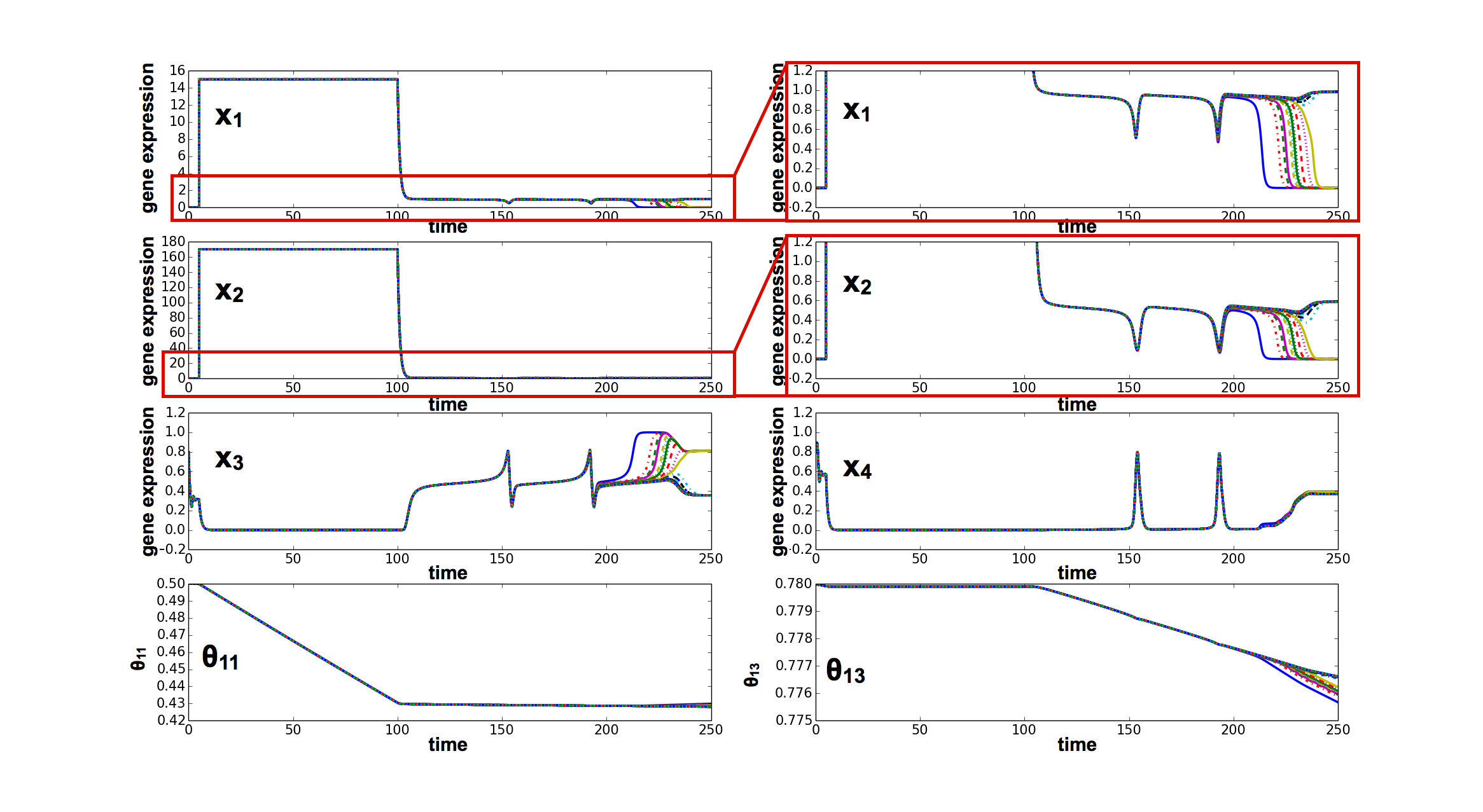}\\
{\bf S4 Fig. Time series of gene expression in the reprogramming simulation by inducing two factors.} Specifically, time series of gene expression for $x_1, x_2, x_3, x_4$, and the epigenetic variables $\theta_{11}$ and $\theta_{13}$ are shown. The initial condition was as follows: all cells were in the differentiated state, where the epigenetic threshold values were set at $1.0$ for the pluripotent genes $\Theta_{31}, \Theta_{21}$, and $\Theta_{42}$, and lowered for the differentiation regulators to $\Theta_{13}=0.78$, $\Theta_{34}=0.5$, $\Theta_{43}=0.3$. The value of the auto-regulator $\Theta_{11}$ was set at $0.50$. In differentiated cells, genes $x_1$ and $x_2$ were overexpressed for a long period. The epigenetic threshold $\theta_{ij}$ decreased with the overexpression of these genes, and the gene expression restarted oscillation. Later, a few cells differentiated again; thus, cells were reprogrammed.

\subsection*{S5 Fig}
\label{FigS4}
\includegraphics*[width=15cm, bb=0 0 2331 1277]{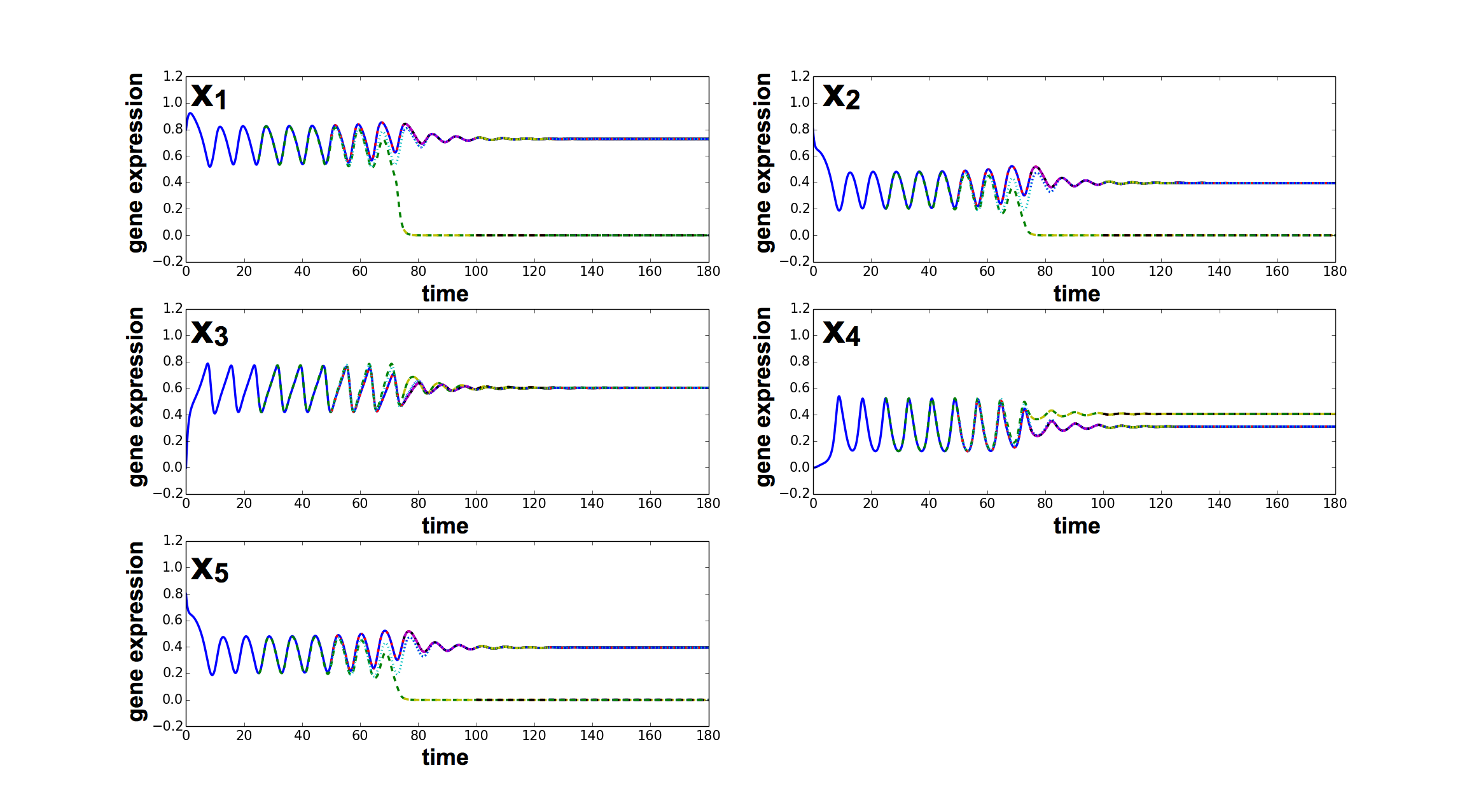}\\
{\bf S5 Fig. Time series of gene expression in the five-gene model.} Time series of gene expression levels for $x_1$, $x_2$, $x_3$, $x_4$, and $x_5$. Expression levels of cells are plotted according to color, but most colors are overlaid and, therefore, difficult to discern. Here, the following parameter set was used: $K_{13}=0.80, K_{34}=0.45, K_{43}=0.45, K_{15}=0.14 K_{31}=0.94, K_{21}=0.81, K_{51}=0.81$, and $K_{42}=0.30$. Gene expression levels initially showed oscillation, and then they were desynchronized. Ultimately, this model showed differentiation, as observed in the four-gene model.

\subsection*{S6 Fig}
\label{FigS5}
\includegraphics*[width=15cm, bb=0 0 2331 1277]{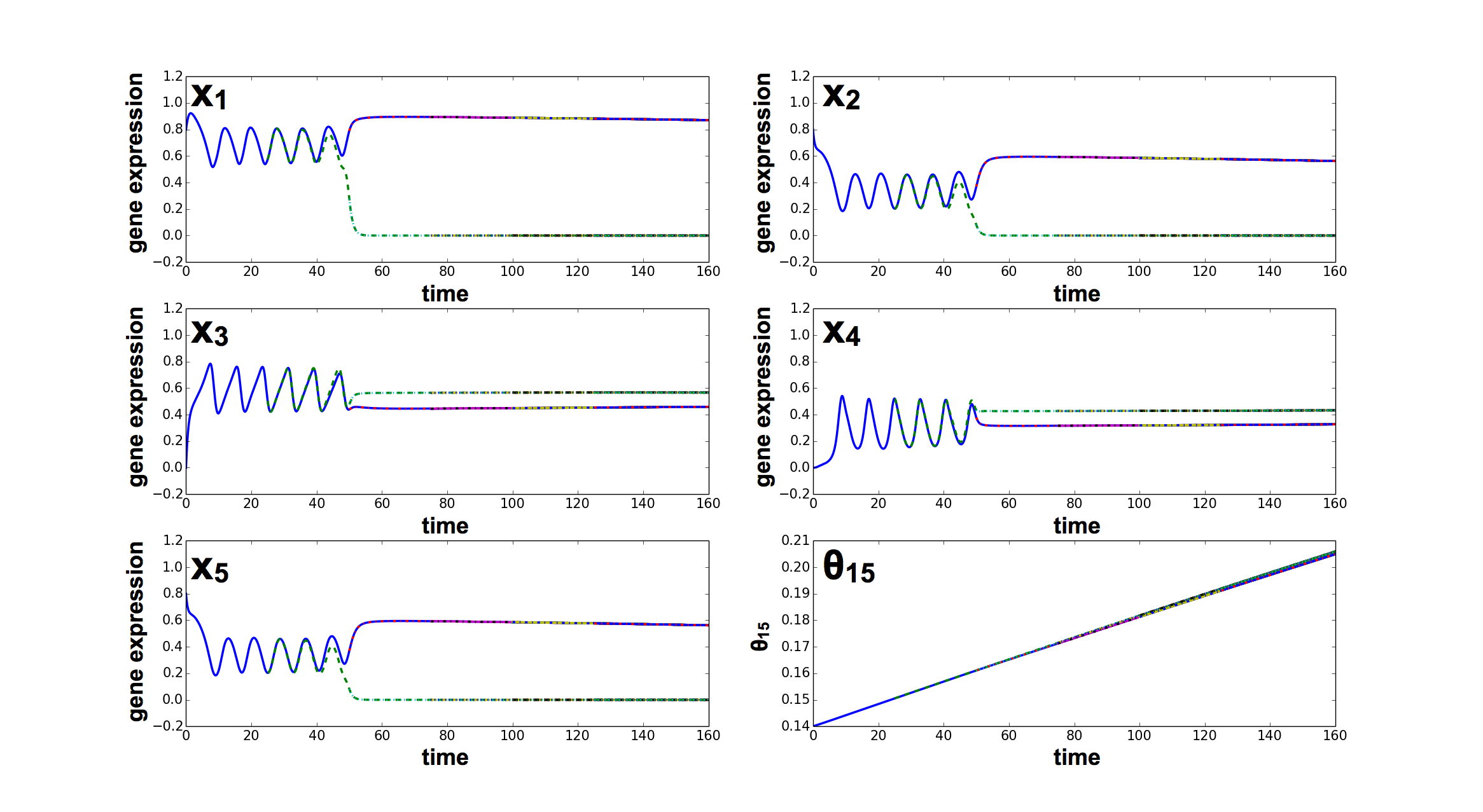}\\
{\bf S6 Fig. Time series of gene expression in the five-gene model with the epigenetic parameter.} Time series of gene expression levels for $x_1$, $x_2$, $x_3$, $x_4$, $x_5$, and the epigenetic threshold variables $\theta_{15}(t)$. Here, we used parameters of $\Theta_{ij}$ as follows: $\Theta_{13}=\Theta_{34}=\Theta_{43}=0.65, \Theta_{15}=\Theta_{31}=\Theta_{21}=\Theta_{51}=\Theta_{42}=1.0$. Initially, gene expression oscillated and gradually desynchronized with cell division. Ultimately, cells fell into a fixed point $x_1 \sim 1$ or $x_1 \sim 0$ because of epigenetic fixation.

\subsection*{S7 Fig}
\label{FigS6}
\includegraphics*[width=6cm, bb=0 0 545 355]{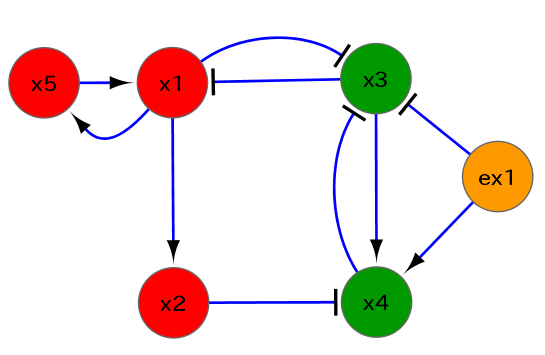}\\
{\bf S7 Fig. The gene regulatory network in the reprogramming simulation with the five-gene model.} The GRN for reprogramming in the five-gene model and an external factor for reprogramming. Arrow headed and T-headed lines represent positive and negative regulation, respectively. The pluripotent genes $x_1$, $x_2$, and $x_5$ were overexpressed, and the external stimulus $ex1$ was added to inhibit $x_3$. Cells were reprogrammed by the induction of these factors, by which cells restarted oscillation. These inducing factors corresponded to the Yamanaka factors (\textit{Oct4, Sox2, Klf4}, and \textit{Myc}).

\end{document}